\documentclass[prb,aps,twocolumn,draft,showpacs]{revtex4}

\usepackage{graphicx}
 \usepackage{amsmath}
 \usepackage{graphpap}
 \usepackage{rotating}

\newcommand{\sigmaS}{\sigma_s}
\newcommand{\drho}{\delta\rho}
\newcommand{\e}{\epsilon}
\newcommand{\la}{\lambda}
\newcommand{\La}{{\bf\Lambda}}
\newcommand{\Labar}{\bar{\bf\Lambda}}
\newcommand{\dLa}{\delta\La}
\newcommand{\weyl}{{\mbox{\tiny Weyl}}}
\newcommand{\cO}{{\cal O}}
\newcommand{\cP}{{\cal P}}
\newcommand{\cN}{{\cal N}}
\newcommand{\cV}{{\cal V}}
\newcommand{\cA}{{\cal A}}
\newcommand{\cS}{{\cal S}}
\newcommand{\cC}{{\cal C}}
\newcommand{\cJ}{{\cal J}}
\newcommand{\cH}{{\cal H}}
\newcommand{\cB}{{\cal B}}
\newcommand{\cR}{{\cal R}}
\newcommand{\cI}{{\cal I}}

\newcommand{\bbmM}{\mathrm{I\!\bf M}}
\newcommand{\unitMat}{\openone}

\newcommand{\del}[1]{\frac{\partial}{\partial #1}}
\newcommand{\Del}[2]{\frac{\partial #1}{\partial #2}}
\newcommand{\ddel}[1]{\frac{\partial^2}{\partial {#1}^2}}

\newcommand{\mel}[4][{}]{\left\langle #2\left\vert\,#3\right\vert\,#4\right\rangle_{#1}}
\newcommand{\scp}[3][{}]{\left\langle #2\left\vert\,#3\right.\right\rangle_{#1}}

\begin{document}

\title{Stability and Symmetry Breaking in Metal Nanowires}

\author{D.~F.~Urban$^1$, J.~B\"urki$^{2}$,
C.~A.~Stafford$^2$, and Hermann~Grabert$^1$}

\affiliation{${}^1$Physikalisches Institut, Albert-Ludwigs-Universit\"at, 
	D-79104 Freiburg, Germany \\ 
	${}^2$Department of Physics, University of Arizona, Tucson, AZ 85721}


\date{August 16, 2006}

\begin{abstract} 
A general linear stability analysis of simple metal nanowires is presented using a 
continuum approach which correctly accounts for material-specific surface properties 
and electronic quantum-size effects. The competition between surface tension and 
electron-shell effects leads to a complex landscape of stable structures as a 
function of diameter, cross section, and temperature. By considering arbitrary 
symmetry-breaking deformations, it is shown that the cylinder is the only generically 
stable structure. Nevertheless, a plethora of structures with broken axial symmetry 
is found at low conductance values, including wires with quadrupolar, hexapolar and 
octupolar cross sections. These non-integrable shapes are compared to previous 
results on elliptical cross sections, and their material-dependent relative stability 
is discussed.
\end{abstract}

\maketitle \vskip2pc

\section{Introduction}\label{sec:intro}

Free-standing metal nanowires, suspended from electrical contacts at their ends, have 
been fabricated by a number of different techniques.\cite{Agrait03} Metal wires down 
to a single atom thick were extruded using a scanning-tunneling microscope tip. 
\cite{Rubio96,Untiedt97} Metal nanobridges were shown to ``self-assemble'' under 
electron-beam irradiation of thin metal films,\cite{Kondo97,Rodrigues00,Kondo00} 
leading to nearly perfect cylinders down to four atoms in diameter, with lengths up 
to fifteen nanometers. Systematic studies of nanowire properties for a variety of 
materials were carried out using the mechanically-controllable break junction 
technique.\cite{Yanson99,Yanson00,Yanson01,Diaz03,Mares04,Mares05} As the ultimate 
nanoscale conductors, metal nanowires are of great interest for nanotechnology.

A remarkable feature of metal nanowires is that they are stable at all. Most atoms in 
such a thin wire are at the surface, with small coordination numbers, so that {\em 
surface effects} play a key role in their energetics. Indeed, macroscopic arguments 
comparing the surface-induced stress to the yield strength indicate a minimum radius 
for solidity of order ten nanometers.\cite{Zhang03} Below this critical radius, 
plastic flow would lead to a Rayleigh instability, \cite{Chandrasekhar81} absent some 
other stabilizing mechanism.

A series of experiments on alkali metal nanocontacts\cite{Yanson99,Yanson00,Yanson01} 
identified {\em electron-shell effects} as another key mechanism influencing nanowire 
stability. Energetically-favorable structures were revealed as peaks in conductance 
histograms, periodic in the nanowire radius, analogous to the electron-shell 
structure previously observed in metal clusters.\cite{deHeer93} A supershell 
structure was also observed,\cite{Yanson00} in the form of a periodic modulation of 
the peak heights. Recently, electron-shell structure has also been observed for the 
noble metals gold,\cite{Diaz03,Mares04} copper,\cite{Mares05} and 
silver.\cite{Mares05}

A theoretical analysis using the nanoscale free-electron 
model\cite{Stafford97a,Buerki05b} (NFEM) found that nanowire stability is determined 
by the competition of these two key factors, surface tension and electron-shell 
effects. Both linear\cite{Kassubek01,Zhang03,Urban03} and 
nonlinear\cite{Buerki03,Buerki05} stability of axially symmetric nanowires were 
investigated. It was found that the surface-tension driven instability can be 
completely suppressed in the vicinity of certain ``magic radii.''

It is well known in the physics of crystals and molecules that a Jahn-Teller 
deformation breaking the symmetry of the system can be energetically favorable. In 
metal clusters, Jahn-Teller deformations are also very 
common,\cite{bulgac93,Schmidt99} and most of the observed structures show a broken 
spherical symmetry. By analogy, it is natural to assume that for nanowires, too, a 
breaking of axial symmetry can be energetically favorable, and lead to more stable 
deformed geometries.

Recently, as a first step towards a stability analysis of wires with a more general 
cross section, elliptical wires were examined within the NFEM.\cite{Urban04} The 
sequence of stable cylindrical and elliptical nanowires allows for a consistent 
interpretation of experimental conductance histograms for alkali metals, including 
both the electronic shell and supershell structures.\cite{Urban04b}

Note that while the experimental manifestations of electron-shell structure are 
similar in metal clusters and nanowires, the Jahn-Teller effect plays out quite 
differently in these two systems. The fundamental difference is that surface effects 
tend to stabilize clusters, while they are the source of the Rayleigh instability in 
nanowires.\cite{Kassubek01} Therefore, Jahn-Teller deformations of clusters are very 
common and typically rather small,\cite{bulgac93,Schmidt99} while they only occur for 
a minority of stable nanowires and can be rather large.

Cylindrical and elliptical nanowires are special, in the sense that the classical 
electron dynamics in these structures is {\em integrable}. An open question is 
whether integrability per se plays a special role in nanowire stability.  On the one 
hand, the shell effect is enhanced in integrable systems,\cite{Brack97} which argues 
in favor of integrability.  On the other hand, the Jahn-Teller effect is driven by 
the lifting of degeneracy due to symmetry breaking, and the degeneracy of states with 
angular momentum $\pm \mu \hbar$ about the axis of symmetry is broken to leading 
order by a perturbation of the radius $\delta r(\phi) \propto \cos (2\mu\phi)$, where 
$\phi$ is the azimuthal angle, which renders the dynamics {\em chaotic}. Furthermore, 
for the case $\mu=1$ (quadrupolar cross section), the surface-energy cost of the 
perturbation is somewhat {\em less} than that of an elliptical deformation with the 
same aspect ratio.

In this article, we perform a general linear stability analysis for straight metal 
nanowires, including arbitrarily shaped, non-integrable cross sections. By 
considering all symmetry-breaking deformations, it is rigorously shown that the 
cylinder is the only generically stable structure. We derive the complete stability 
diagram for cylinders, and discuss why axial symmetry is special---in the sense that 
75\% of the experimentally observed alkali metal nanowires are indeed cylindrical, 
and that proportion goes up with increasing conductance.

Nevertheless, a wide range of structures with broken axial symmetry is found at low 
conductance values, most of which show a reduced relative stability compared to the 
axially symmetric ones. We examine wires with quadrupolar, hexapolar and octupolar 
cross sections, and compare these non-integrable shapes to the previous results on 
elliptical cross sections. Their relative stability and their degree of deformation 
are material-dependent, and we examine these properties for the alkali and noble 
metals.

This paper is organized as follows: The assumptions and features of the nanoscale 
free-electron model are summarized in Sec.\ \ref{sec:model}, followed by the 
presentation of the linear stability analysis in Sec.\ \ref{sec:StabAna}. Cylindrical 
wires are examined in Sec.\ \ref{sec:Cylinders}, whereas Sec.\ \ref{sec:JahnTeller} 
considers non-integrable cross sections. The material dependence of our results is 
discussed in Sec.\ \ref{sec:materialdependence}. Finally, we summarize and discuss 
our results in Sec.\ \ref{sec:discussion}. The Appendix provides additional technical 
details of our calculations.

\section{Nanoscale Free Electron Model}\label{sec:model}

Guided by the importance of conduction electrons in the cohesion of metals, and by 
the success of the jellium model in describing metal clusters,\cite{deHeer93,Brack93} 
the nanoscale free-electron model \cite{Stafford97a,Buerki05b} replaces the metal 
ions by a uniform, positively charged background that provides a confining potential 
for the electrons. The electron motion is free along the wire, and confined in the 
transverse directions. Due to the excellent screening\cite{Kassubek99,Zhang05} in 
metal wires with $G>G_0$, where $G_0=2e^2/h$ is the conductance quantum, 
electron-electron interactions can in most cases be neglected. The surface properties 
of various metals can be fit by using appropriate surface boundary 
conditions.\cite{Garcia-martin96,Urban04}

The NFEM is especially suitable for alkali metals, but is also adequate to describe 
shell effects due to the conduction-band $s$-electrons in other monovalent metals, 
such as noble metals, and in particular, gold. The experimental observation of a 
crossover from atomic-shell to electron-shell effects with decreasing radius in both 
metal clusters\cite{Martin96} and nanowires\cite{Yanson01} justifies {\it a 
posteriori} the use of the NFEM in the later regime.

Since a nanowire connecting two macroscopic electrodes is an open quantum system, the 
Schr\"odinger equation is most naturally formulated as a scattering problem. The 
fundamental theoretical quantity is the scattering matrix $S(E)$ connecting incoming 
and outgoing asymptotic states of conduction electrons in the electrodes. 
Thermodynamic properties can be expressed in terms of the scattering matrix through 
the electronic density of states\cite{Dashen69}
\begin{eqnarray}
\label{gl.DausS}
        D(E) &=&
    \frac{1}{2\pi i}\;\mbox{Tr}\left\{
    S^{\dagger}(E)\frac{\partial S}{\partial E} -
        \frac{\partial S^{\dagger}}{\partial E}S(E)\right\},
\end{eqnarray}
from which the relevant thermodynamic potential for an open system, namely the grand 
canonical potential $\Omega$, is obtained as
\begin{equation}
\label{gl.OmegaVonD}
    \Omega=-\frac{1}{\beta} \int \!dE\; D(E) \;
    \ln\!\left[1+e^{-\beta(E-\mu)}
        \right].
\end{equation}
Here $\beta=(k_BT)^{-1}$ is the inverse temperature and $\mu$ is the electron 
chemical potential, specified by the macroscopic electrodes. Eqs.~(\ref{gl.DausS}) 
and (\ref{gl.OmegaVonD}) include a factor of 2 for spin degeneracy.

Any extensive thermodynamic quantity can be expressed as the sum of a Weyl expansion, 
which depends on geometrical quantities such as the system volume $\cV$, surface area 
$\cS$, and integrated mean curvature $\cC$, and an oscillatory shell-correction due 
to quantum-size effects.\cite{Brack97} In particular, the grand canonical potential 
(\ref{gl.OmegaVonD}) can be written as
\begin{equation}
    \label{eq:OmegaWeyl}
    \Omega = -\omega\,\cV
    +\sigmaS\, {\cal S}
    -\gamma_s\,\cC
    +\delta\Omega,
\end{equation}
where the energy density $-\omega$, surface tension coefficient $\sigmaS$, and 
curvature energy $-\gamma_s$ are, in general, material- and temperature-dependent 
coefficients. On the other hand, the shell correction $\delta\Omega$ can be shown, 
based on very general arguments,\cite{Strutinsky68,Stafford99,Zhang05} to be a 
single-particle effect, which is well described by the NFEM.

The ionic degrees of freedom in the NFEM are modeled as an incompressible 
fluid.\cite{Zhang03,Buerki03,Buerki05} This takes into account, to lowest order, the 
hard-core repulsion of the core electrons as well as the exchange energy of the 
conduction electrons. Any atom-conserving deformation of the structure is therefore 
subject to a constraint of the form
\begin{equation}
\label{eq:constraint}
  {\cal N} \equiv k_F^3{\cal V}
    - \eta\, k_F^2{\cal S}
    + \xi\,k_F{\cal C}\, = \rm{const},
\end{equation}
where $k_F$ is the Fermi wavevector. The parameters $\eta$ and $\xi$ can be adjusted 
so as to fix the values of the effective surface tension and curvature energy in the 
free-electron model to the material-specific values, as discussed in detail in Sec.\ 
\ref{sec:materialdependence}. However, the shell correction $\delta\Omega$ is (to 
leading order) independent\cite{Stafford99,Urban04} of $\eta$ and $\xi$.

In this article, the shell correction $\delta\Omega$ is calculated 
quantum-mechanically, from Eqs.\ (\ref{gl.DausS}) and (\ref{gl.OmegaVonD}).  
Nonetheless, in interpreting our quantum-mechanical results, it is useful to appeal 
to semiclassical arguments,\cite{Gutzwiller90,Brack97} according to which the shell 
correction can be computed by a Gutzwiller-type sum\cite{Gutzwiller71} over classical 
periodic orbits, that takes the form of an integral over the length $L$ of the 
nanowire\cite{Buerki03}
\begin{equation}\label{eq:TraceFormula}
 \delta\Omega= \frac{E_F}{\lambda_F}\int_0^L dz \sum_{po}A_{po}(z,T)\cos(k_FL_{po}(z)+\theta_{po}),
\end{equation}
where $E_F$ and $\lambda_F$ are the Fermi energy and Fermi wavelength, respectively. 
Here the sum includes all classical periodic orbits lying in the plane of the wire's 
cross section, $L_{po}$ is the orbit length, and $\theta_{po}$ is an orbit-dependent 
phase. Semiclassically,\cite{Brack97} the temperature dependence of $\delta\Omega$ 
appears only in the dimensionless amplitude $A_{po}$, which also depends on the 
stability, symmetry, and length of the periodic orbit. The orbit amplitude has the 
form\cite{Brack97}
\begin{equation}
\label{eq:amplitude}
    A_{po} \sim (k_F L_{po})^{-\nu_{po}} \frac{\tau_{po}}{\sinh \tau_{po}}, 
    \;\;\; \tau_{po} \equiv \frac{\pi k_F L_{po} T}{2 T_F},
\end{equation}
where $T_F=E_F/k_B$ is the Fermi temperature, and $\nu_{po}=1$ for a continuous 
family of degenerate orbits, which arise when the cross section has the form of an 
integrable billiard, such as a disk or an ellipse, and $\nu_{po}=3/2$ for an isolated 
orbit in a wire with a cross section whose dynamics is fully chaotic (see Ref.\ 
\onlinecite{Stafford01} for details).

It is useful to define the r.m.s radius $\rho$ of the cross section, which is related 
to the cross-sectional area by $\cA = \pi \rho^2$.  Then $L_{po} \propto \rho$, so 
that $\delta\Omega \sim \rho^{-1}$ for an integrable cross section, and decays 
exponentially for temperatures exceeding the radius-dependent characteristic 
temperature $T_\rho \equiv T_F/k_F\rho$. As shown below, these semiclassical 
estimates of the temperature- and size-dependence of $\delta\Omega$ are in accord 
with our quantum-mechanical results.

\section{Linear Stability Analysis}\label{sec:StabAna}

Let us consider a wire with uniform cross section aligned along the $z$-axis. Its 
surface is given by the radius function $r=R(\varphi;\rho,\La)$ in cylindrical 
coordinates $r,\,\varphi,\,z$, where the parameters $\{\rho,\La\}$ describe the cross 
section geometry, chosen such that the cross-sectional area is given by ${\cal 
A}=\pi\rho^2$, whereas the shape is determined by a set of dimensionless parameters, 
composing the vector $\La$. Without loss of generality, we can associate $\La=0$ with 
a cylindrical wire.

A small $z$-dependent perturbation of a wire of length $L$ and initial cross section 
$(\bar{\rho},\Labar)$ can be written in terms of a Fourier series as
\begin{eqnarray}
  \label{eq:perturbation}
     \rho(z)  &=&\bar{\rho} + \e\,\drho(z) \;=\; \bar{\rho} + \e\sum_q\rho_q\,e^{i q z},
\nonumber\\
     \La(z) &=&\Labar + \e\,\dLa(z)\;=\;\Labar + \e\sum_q\La_q\,e^{i q z},
\end{eqnarray}
where the dimensionless small parameter $\e$ sets the size of the perturbation. 
Assuming periodic boundary conditions, the perturbation wave vectors $q$ must be 
integer multiples of $2\pi/L$. In order to ensure that $\rho(z)$ and $\La(z)$ are 
real, we have $\rho_{-q}=\rho^*_q$ and $\La_{-q}=\La^*_q$.

The structural stability of metal nanowires is governed by the response to 
long-wavelength perturbations,\cite{Kassubek01,Zhang03,Urban04,Buerki05b} while the 
response to short-wavelength perturbations controls surface quantum fluctuations in 
long wires.\cite{Urban03,Urban06} Since we consider free electrons, the stationary 
Schr\"odinger equation is given by $-\triangle\Psi=(2m_eE/\hbar^2)\Psi$, where 
$\triangle$ is the Laplace operator in cylindrical coordinates. In the 
long-wavelength limit, the Schr\"odinger equation may be solved in the adiabatic 
approximation, for which transverse and longitudinal motions decouple. Therefore, we 
use the ansatz $\Psi(r,\varphi,z)=\chi(r,\varphi;z)\Phi(z)$, and neglect all 
$z$-derivatives of the transverse wavefunction $\chi$, so that the Schr\"odinger 
equation becomes
\begin{eqnarray}
\label{eq:Seq:transverse}
    \left(\ddel{r}\! + \!\frac{1}{r}\del{r}
          \! + \!\frac{1}{r^2}\ddel{\varphi}\! + \!\frac{2m_e}{\hbar^2}E_n(z)\!\right)
      \chi_n(r,\varphi;z)&=&0,\quad
  \\
\label{eq:Seq:longitudinal}
    \left(\ddel{z}+\frac{2m_e}{\hbar^2}\left(E-E_n(z)\right)\right)\Phi_n(z)&=&0.\quad
\end{eqnarray}
First, the transverse problem (\ref{eq:Seq:transverse}) is solved at fixed $z$ in 
order to determine the transverse eigenenergies $E_{n}(z)$. With our choice of cross 
section parametrization, their dependence on geometry can be written as
\begin{eqnarray}
    E_{n}(\rho,\La)&=&\frac{\hbar^2}{2m_e}\left(\frac{\gamma_{n}(\La)}{\rho}\right)^2,
\end{eqnarray}
where the function $\gamma_{n}(\La)$ remains to be determined. For an axisymmetric 
wire, the $\gamma_{n}$ are given by the roots of the Bessel functions, whereas for an 
elliptical cross section, the $\gamma_{n}$ are given by the roots of the modified 
Mathieu functions.\cite{Urban04} In general, the shape-dependent $\gamma_{n}(\La)$ 
are determined numerically (see Appendix for details).

We are then left with a series of effective one-dimensional scattering problems (Eq.\ 
\ref{eq:Seq:longitudinal}) for the longitudinal wave functions $\Phi_n(z)$, in which 
the transverse eigenenergies $E_{n}\big(\rho(z),\La(z)\big)$ act as additional 
potentials for the motion along the wire. These scattering problems can be solved 
using the WKB approximation\cite{Stafford97a}: The grand canonical potential is given 
by
\begin{equation}
\label{eq:omegaT}
  \Omega[T;\rho,\La] = \int_0^\infty dE\left(-\frac{\partial f}{\partial
  E}\right)\Xi[E;\rho,\La],
\end{equation}
where $f=(1+\exp[(E-\mu)/k_B T])^{-1}$ is the Fermi function at temperature $T$ and 
chemical potential $\mu\approx E_F$, and $\Xi$ is given by
\begin{equation}
\label{eq:omega}
    \Xi[E;\rho,\La] =
    -\frac{8E_F}{3\lambda_F}
    \int_0^L\!\!\text{d}z
    \sum_{n}\left(\frac{E-E_{n}(\rho,\La)}{E_F}\right)^{\!3/2}\!\!\!\!\!.
\end{equation}
The sum runs over all open channels ${n}$, for which $E_{n}(\rho,\La)<E$.

The energetic cost of a small deformation of the wire can be calculated by expanding 
Eq.~(\ref{eq:omegaT}) as a series in the parameter $\e$,
\begin{equation}
\label{eq:omega.expand}
    \Omega=\Omega^{(0)}+\e\,\Omega^{(1)}+\e^2\,\Omega^{(2)}+{\cal O}(\e^3).
\end{equation}
A nanowire with initial cross section $(\bar{\rho},\Labar)$ is energetically stable 
at temperature $T$ if and only if $\Omega^{(1)}(\bar{\rho},\Labar)=0$ and 
$\Omega^{(2)}(\bar{\rho},\Labar)>0$ for every possible deformation $(\drho,\dLa)$ 
satisfying the constraint (\ref{eq:constraint}).

A straightforward expansion of Eq.\ (\ref{eq:omegaT}) at $T=0$ yields
\begin{equation}
    \frac{\Omega^{(1)}}{L/\la_F} 
    = 4\sum_{n} \sqrt{\frac{E_F-\bar{E}_{n}}{E_F}}
      \left(\La_0\!\cdot\!\nabla_{\!\La}\bar{E}_{n}-2\bar{E}_{n}\frac{\rho_0}{\bar{\rho}}\right),
\label{eq:omega(1)} 
\end{equation}
\begin{equation}
\label{eq:omega12:m}
  \frac{\Omega^{(2)}}{L/\la_F} = E_F
  \sum_q
    \binom{\rho_q/\bar{\rho}}{\La_q}^{\!\!\dagger} \!\!
    \begin{pmatrix} A_{\rho\rho} & \!A_{\rho\La} \\ A_{\La\rho} &
\!A_{\La\La}\end{pmatrix}\!\!    \binom{\rho_q/\bar{\rho}}{\La_q},
\end{equation}
where the elements of the matrix $A$ in Eq.\ (\ref{eq:omega12:m}) are given by
\begin{eqnarray}
\label{eq:stabcoef:m}
  A_{\rho\rho}&\!\!=\!\!&\sum_{n} \frac{4\bar{E}_{n}}{E_F^{3/2}}\left[
    3\sqrt{E_F\!-\!\bar{E}_{n}}-\frac{\bar{E}_{n}}{\sqrt{E_F\!-\!\bar{E}_{n}}}
    \right],
\nonumber 
\\
  A_{\La\rho}&\!\!=\!\!&-\sum_{n}\frac{4\bar{E}^\prime_n}{E_F^{3/2}}
    \left[
      \sqrt{E_F\!-\!\bar{E}_{n}}
      -\frac{\bar{E}_n}{2\sqrt{E_F\!-\!\bar{E}_n}}
    \right]\!,
\\
  A_{\La\La}&\!\!=\!\!&\sum_{n}\frac1{E_F^{3/2}}\left[
    2\bar{E}^{\prime\prime}_n\sqrt{E_F\!-\!\bar{E}_{n}}
    - \frac{\bar{E}^{\prime}_n\cdot(\bar{E}^{\prime}_n)^{\dagger}}
        {\sqrt{E_F\!-\!\bar{E}_{n}}}
    \right]\!. 
\nonumber 
\end{eqnarray}
Here $\bar{E}_n=E_n(\bar{\rho},\Labar)$, $\bar{E}^{\prime}_n=\nabla_\La 
E_n|_{\bar{\rho},\Labar}$ denotes the gradient of $E_{n}$ with respect to $\La$, and 
$(\bar{E}^{\prime\prime}_n)_{ij}=\left.\partial^2E_{n}/\partial\Lambda_i\partial\Lambda_j\right|_{\bar{\rho},\Labar}$ 
is the matrix of second derivatives, all evaluated at $(\bar{\rho},\Labar)$.

The number of independent Fourier coefficients in Eq.\ (\ref{eq:perturbation}) is 
restricted through the constraint (\ref{eq:constraint}) on allowed deformations. 
Hence, after evaluating the change of the geometric quantities $\cV$, $\cS$, and 
$\cC$ due to the deformation, we can use Eq.\ (\ref{eq:constraint}) to express 
$\rho_0$ in terms of the other Fourier coefficients. This yields an expansion 
$\rho_0=\rho_0^{(0)}+\e\,\rho_0^{(1)}+\cO(\e^2)$, with
\begin{eqnarray}
    \label{eq:sigma0}
  \frac{\rho_0^{(0)}}{\bar{\rho}} &\!=\!&
  \frac{\eta}{k_F\bar{\rho}-\eta\bar{P}}\;\bar{P}^{\prime}\cdot\La_0\,,
\\
  \frac{\rho_0^{(1)}}{\bar{\rho}} &\!=\!&\frac{1}{2(k_F\bar{\rho}\!-\!\eta
    \bar{P})}\sum_q \binom{\rho_q/\bar{\rho}}{\La_q}^{\!\!\dagger} \!\!
    \begin{pmatrix}
      -k_F\bar{\rho}     &    \eta\bar{P}' \\
      \eta(\bar{P}')^T   &    \eta\bar{P}''
    \end{pmatrix}
    \!\!\binom{\rho_q/\bar{\rho}}{\La_q},\nonumber
\end{eqnarray}
where the function $\bar{P}=P(\Labar)$ is related to the perimeter $\cP$ by 
$\cP(\bar{\rho},\Labar)=2\pi\bar{\rho}\bar{P}$, the first derivative 
$\bar{P}'=\nabla_{\!\La}P|_{\Labar}$ is a vector, and the second derivative 
$\bar{P}''=\partial^2P/\partial\Lambda_i\partial\Lambda_j|_{\Labar}$ is a matrix. 
Note that in the long-wavelength limit considered here, Eq.\ (\ref{eq:sigma0}) is 
independent of the parameter $\xi$ appearing in the constraint (\ref{eq:constraint}), 
so that our results are independent of the value of $\xi$. Inserting the expansion 
for $\rho_0$ in Eq.\ (\ref{eq:omega(1)}), the first-order change of the energy under 
a constant ${\cal N}$ perturbation is given by
\begin{eqnarray}
\label{eq:omega1N:m}
    \frac{\la_F}{L}\left.\Omega^{(1)}\right|_\cN &=&\frac{4}{\sqrt{E_F}}
    \sum_{n}\sqrt{E_F-\bar{E}_{n}} \\ \nonumber
    & & \times \left(\bar{E}_n^{\prime} 
    -\frac{2\eta\bar{E}_{n}}{k_F\bar{\rho}-\eta\bar{P}}\;\bar{P}^{\prime}
    \right)\cdot\La_0,
\end{eqnarray}
and the second-order term is given by Eq.\ (\ref{eq:omega12:m}), with the matrix $A$ 
replaced by
\begin{eqnarray}
\label{eq:stabcoefN:m}
  \tilde{A}_{\rho\rho}&=& A_{\rho\rho} +
    \frac{4 k_F\bar{\rho}}{k_F\bar{\rho}-\eta\bar{P}}
    \sum_{n} \frac{\bar{E}_{n} \sqrt{E_F\!-\!\bar{E}_{n}}}{E_F^{3/2}},
    \nonumber
 \\
  \tilde{A}_{\La\rho}&=& A_{\La\rho}
- \frac{4\eta P^{\prime}} {k_F\bar{\rho}-\eta\bar{P}}
  \sum_{n}  \sqrt{\frac{E_F\!-\!\bar{E}_{n}}{E_F}},
 \\
  \tilde{A}_{\La\La}&=& A_{\La\La}
    -\frac{4\eta\bar{P}^{\prime\prime}}{k_F\bar{\rho}-\eta\bar{P}}
    \sum_{n}\sqrt{\frac{E_F\!-\!\bar{E}_{n}}{E_F}}. \nonumber
\end{eqnarray}
The stability condition, $\left.\Omega^{(2)}(\bar{\rho},\Labar)\right|_\cN>0$, 
requires that the {\em stability matrix} $\tilde{A}$ be positive definite. The 
results at finite temperature are obtained in a similar fashion, by integrating Eq.\ 
(\ref{eq:omegaT}) numerically.

\section{General stability of cylinders}\label{sec:Cylinders}

\begin{figure*}[bt]        
\newlength\figonewidth\setlength\figonewidth{12.8cm}
    \begin{minipage}[c]{\figonewidth}\hspace*{-6mm}
       \includegraphics[width=\figonewidth,clip=,draft=false]{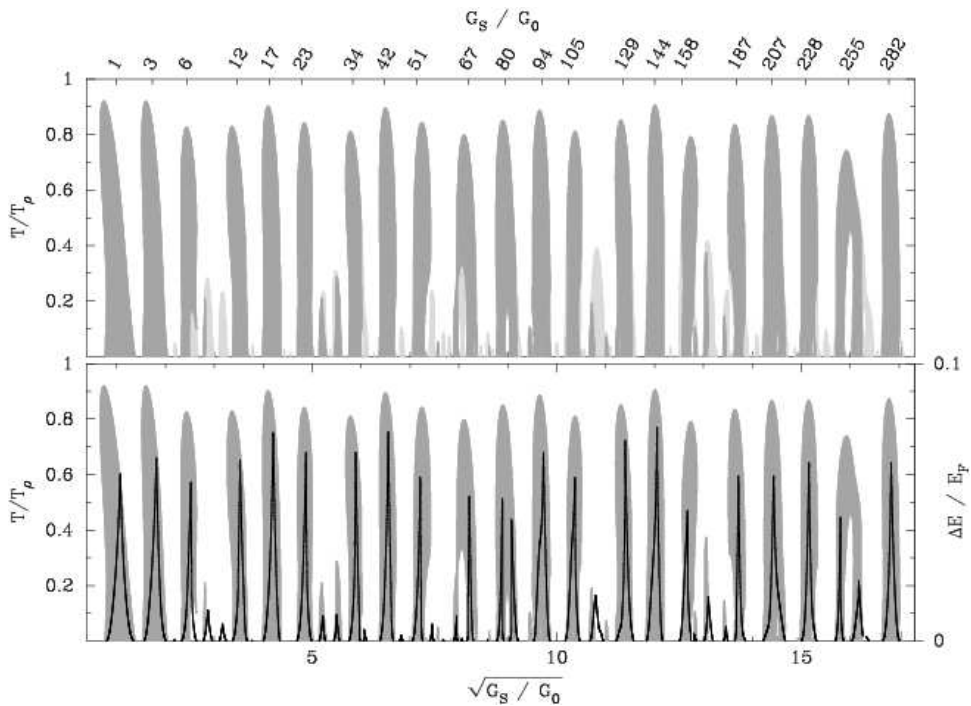}
    \end{minipage}\
    \begin{minipage}[c]{4.5cm}
      \caption[]{
      Stability of metal nanocylinders versus electrical conductance and
      temperature.  Dark gray areas indicate
      stability with respect to arbitrary small deformations.
      Temperature is displayed in units of $T_{\rho}=T_F/k_F\rho$ (see text).
      The surface tension was taken as $0.22\,\rm{N/m}$, corresponding to Na.\cite{Perdew91}
      For comparison, the upper panel also shows wires that
      are stable towards axisymmetric deformations only (light gray areas), but which are
      unstable when allowing symmetry breaking deformations.
      The lower panel includes a plot of the activation energy
      $\Delta E$ (solid line), reproduced from
      Ref.\ \onlinecite{Buerki05}, which determines the nanowire lifetime
      $\tau=\tau_0\exp(\Delta E/k_BT)$.
      } \label{fig:StabCylinders}
    \end{minipage}\vspace*{-3mm}
\end{figure*}

As a first application of the method presented in the previous sections, we derive 
the complete stability diagram for cylinders, i.e., we determine the radii of 
cylindrical wires that are linearly stable with respect to \emph{arbitrary} small, 
long-wavelength deformations. We choose the following general form of the radius 
function describing the surface of the wire,
\begin{equation}
\label{eq:deformation}
    R(\varphi) = \rho\left( \sqrt{1-\sum_m\frac{\la_m^2}{2}}
      + \sum_m\la_m\cos[m(\varphi\!-\!\varphi_m)]\right).
\end{equation}
The sum runs over a set $\bbmM$ of positive integer values, and the deformation 
parameters $\la_m$ governing deviations from axial symmetry are such that $\sum_m 
\la_m^2<2$. Note that the dipole deformation ($m=1$) corresponds, in leading order, 
to a simple translation, plus higher-order multipole deformations. Therefore we can 
restrict our analysis to $m>1$.

At first sight, considering arbitrary deformations, and therefore theoretically an 
infinite number of perturbation parameters $\{\la_m,\varphi_m\}$ seems a formidable 
task. Fortunately, we find that the stability matrix $\tilde{A}$ for a cylinder is 
diagonal, and therefore the different Fourier contributions of the deformation 
decouple.

This can be seen within perturbation theory for small deviations from axial symmetry, 
derived in the Appendix: for energy levels $n$ that are non-degenerate at $\La=0$, 
i.e.\ levels with orbital quantum number $\mu_n=0$, we obtain
\begin{eqnarray}
\label{eq:DEandDDE:mu0}
    \Del{E_{n}(0)}{\Lambda_i}=0
    &\quad\mbox{and}\quad&
    \frac{\partial^2E_{n}(0)}{\partial\Lambda_i\partial\Lambda_j} \propto\delta_{ij},
    \qquad
\end{eqnarray}
whereas for two energy levels $n$ and $n'$ that are degenerate at $\La=0$, i.e. those 
for which $\mu_n = -\mu_{n'} \neq 0$, we get
\begin{equation}
\label{eq:DE:muNot0}
    \Del{E_{n}(0)}{\Lambda_i}=-\Del{E_{n'}(0)}{\Lambda_i},
\end{equation}
and
\begin{equation}
\label{eq:DDE:muNot0}
    \frac{\partial^2E_{n}(0)}{\partial\Lambda_i\partial\Lambda_j} + 
    \frac{\partial^2E_{n'}(0)}{\partial\Lambda_i\partial\Lambda_j} \propto\delta_{ij}.
\end{equation}
Since the matrix $A$, given by Eq.\ (\ref{eq:stabcoef:m}), includes a sum over all 
open channels, the relations (\ref{eq:DEandDDE:mu0}--\ref{eq:DDE:muNot0}) imply that 
$A$ is diagonal. Furthermore, one can straightforwardly show, by expanding the 
perimeter function $P$ in a series in $\Lambda$, that 
$\bar{P}'=\nabla_{\!\La}P|_{\La=0}=0$ and the matrix $\bar{P}''={\partial^2 
P}/{\partial\Lambda_i\partial\Lambda_j}|_{\La=0}$ is diagonal as well. Therefore, we 
conclude that the stability matrix $\tilde{A}$, including the corrections from the 
constraint (\ref{eq:constraint}), is diagonal.  This also shows that the first order 
correction (\ref{eq:omega1N:m}) to $\Omega$ is identically zero for cylinders. This 
result implies a great simplification of the problem, since it allows to determine 
the stability of cylindrical wires with respect to arbitrary deformations through the 
study of a set of pure $m$-deformations, i.e.\ deformations as given by Eq.\ 
(\ref{eq:deformation}) with only one non-zero $\la_m$.

Fig.\ \ref{fig:StabCylinders} shows the stable cylindrical wires in dark gray as a 
function of temperature. The surface tension was fixed at the value\cite{Perdew91} 
for Na, $\sigma_s=0.22\,\mbox{N/m}$ (i.e.\ $\eta=1.1$). The temperature is given in 
units of the characteristic temperature $T_{\rho}=T_F/k_F\rho$, reflecting the 
temperature dependence of the shell correction (\ref{eq:TraceFormula}) to the wire 
energy: Since stability is determined by a competition between the surface and shell 
contributions,\cite{Kassubek01,Buerki05b} the maximum temperature of linear stability 
scales with $T_\rho$, the surface contribution being only weakly dependent on 
temperature. The $x$-axis is given by the corrected Sharvin conductance
\begin{equation}
    \label{eq:sharvin}
    G_S=G_0\left(\frac{k_F^2\rho^2}{4}-\frac{k_F{\cal{P}}}{4\pi}+\frac{1}{6}\right),
\end{equation}
where ${\cal{P}}$ is the perimeter of the cross section.

The stability diagram was obtained by intersecting a set of individual stability 
diagrams allowing $\cos(m\varphi)$-deformations with $m=2,...,8$. Multipolar 
deformations become increasingly costly with increasing $m$, their surface energy 
scaling as $\rho m^2$, and we find that including $m\geq6$ deformations does not 
further modify the final result.

Figure \ref{fig:StabCylinders} confirms the extraordinary stability of a set of wires 
with so called ``magic radii,'' which were identified in previous studies of 
stability under axisymmetric perturbations 
alone.\cite{Kassubek01,Urban03,Zhang03,Buerki05} They exhibit conductance values 
$G/G_0=$ 1, 3, 6, 12, 17, 23, 34, 42, 51,... However, some wires that are barely 
stable when considering only axisymmetric 
perturbations,\cite{Kassubek01,Zhang03,Urban03} e.g., $G/G_0=$ 5, 10, 14,..., shown 
in light gray in the top panel of Fig.\ \ref{fig:StabCylinders}, are found to be 
unstable when allowing more general, symmetry-breaking deformations.

The heights of the dominant stability peaks in Fig. \ref{fig:StabCylinders} exhibit a 
periodic modulation, with minima occurring near $G_S/G_0=$ 9, 29, 59, 117, 170, and 
255. The positions of the first four minima are in perfect agreement with the 
observed supershell structure in conductance histograms of alkali metal 
nanowires,\cite{Yanson00} while we predict the next two minima of the series. The 
supershell structure is expected to continue at larger conductance, but may be 
obscured by atomic-shell effects in experiments. Interestingly, the nodes of the 
supershell structure, where the shell effect for a cylinder is suppressed, are 
precisely where the most stable deformed nanowires are predicted to occur (see Ref.\ 
\onlinecite{Urban04} and the discussion below). Thus symmetry breaking distortions 
and the supershell effect are inextricably linked.

\begin{figure*}[bt]            
\newlength\figtwowidth\setlength\figtwowidth{13.2cm}
    \begin{minipage}[c]{\figtwowidth}\hspace*{-6mm}
       \includegraphics[width=\figtwowidth,clip=,draft=false]{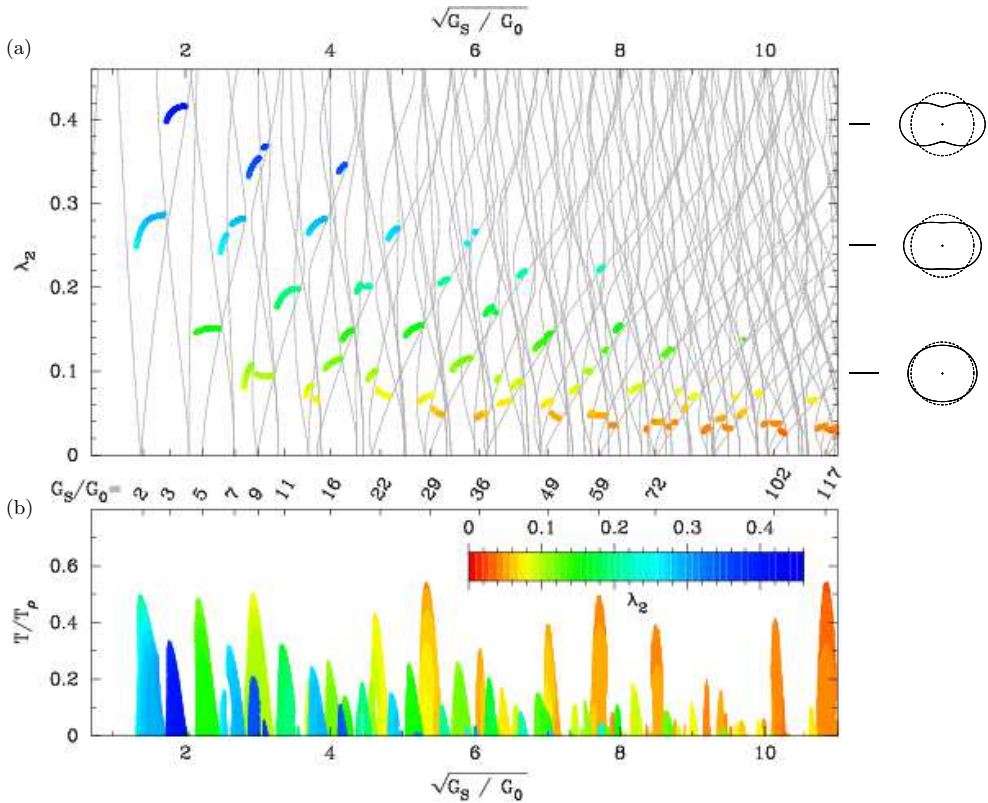}
    \end{minipage}\
    \begin{minipage}[c]{3.8cm}
      \caption[]{(color online)
      (a) Linearly stable quadrupolar ($\lambda_2>0$) Na wires [thick lines] at
      temperature $T=0.03\,T_\rho$, shown in the configuration space of
      the two parameters $\la_2$ and $\rho$ describing the quadrupolar
      shape,
      $R(\varphi)=\rho(\sqrt{1-\lambda_2^2/2}+\lambda_2\cos(2\varphi))$.
      $\rho$ is related to the Sharvin conductance by Eq.\
      (\ref{eq:sharvin}), which gives $\sqrt{G_S/G_0}\approx k_F\rho/2$.
      The thin gray lines show the thresholds for the opening of new
      channels. Some geometries are shown on the right axis,
      illustrating their deviations from axial symmetry; (b) Stability
      diagram for quadrupolar Na wires. The gray(color) scale reflects
      the value of $\lambda_2$ for the stable wires.
      }      \label{fig:Stabdia:m2}
      \end{minipage}
\end{figure*}

Linear stability is a necessary---but not a sufficient---condition for a 
nanostructure to be observed experimentally.  The linearly stable nanocylinders 
revealed in the above analysis are in fact {\em metastable} structures, and an 
analysis of their lifetime has recently been carried out within an axisymmetric 
stochastic field theory.\cite{Buerki05} In the lower panel of Fig.\ 
\ref{fig:StabCylinders}, the zones of linear stability are superimposed on the 
activation energy $\Delta E$ (solid line), calculated in Ref.\ \onlinecite{Buerki05}, 
which determines the nanowire lifetime $\tau$ through the Kramers formula 
$\tau=\tau_0\exp(\Delta E/k_BT)$. As can be seen, there is a strong correlation 
between the size of the activation barriers and the height of the stable fingers in 
the linear stability analysis. This suggests that the linear stability analysis, with 
temperature expressed in units of $T_\rho=T_F/k_F\rho$, provides a good measure of 
the total stability of metal nanowires. In particular, the ``universal'' 
stability\cite{Buerki05} of the most stable cylinders is reproduced, wherein the 
absolute stability of the magic cylinders is essentially independent of radius (aside 
from the small supershell oscillations).

For a given material, the typical value $\Delta E$ for the magic cylinders is 
universal, i.e. it depends only on the surface tension \cite{Buerki05} $\sigmaS$
\begin{equation}
\label{eq:universal}
  \Delta E \approx 0.6 \sqrt{\hbar^2 \sigma_s/m_e},
\end{equation}
where $m_e$ is the electron effective mass, which is of order the free electron mass 
in simple metals. When comparing the \emph{absolute} stability for different 
materials, it is not sufficient to simply compare the heights of the stable fingers 
in Fig.\ \ref{fig:StabCylinders}, which depend only logarithmically on the ratio 
$\sigma_s/E_F k_F^2$ through Eqs.\ (\ref{eq:OmegaWeyl}) and (\ref{eq:TraceFormula}); 
the $\sigmaS^{1/2}$ dependence of the activation barrier, which determines the 
lifetime, must be taken into account explicitly (see Sec.\ 
\ref{sec:materialdependence}).

Since the activation barriers are calculated asymptotically for low 
temperatures,\cite{Buerki05} they exhibit all of the fine structure of the 
low-temperature shell correction.  This fine structure is also seen in the linear 
stability analysis at low temperature (see Fig.\ \ref{fig:StabCylinders}, lower 
panel). However, at higher temperatures, the fine structure in the shell potential is 
smoothed out, as described in Eqs.\ (\ref{eq:TraceFormula}) and (\ref{eq:amplitude}), 
and some of the separate low-temperature stability zones merge into arches at higher 
temperatures. Prominent examples of this phenomenon, the unusual thermodynamics of 
which is discussed in Ref.\ \onlinecite{Zhang03}, occur at $G_S/G_0 \sim 67$, 80, and 
255.

\section{Breaking axial symmetry}\label{sec:JahnTeller}

In a recent paper,\cite{Urban04} the stability of wires with elliptic cross sections 
was determined. Elliptic deformations are special, in that the system remains {\em 
integrable}. Here we focus on wires with a $\cos(m\varphi)$-deformed cross section 
(i.e.\ having $m$-fold symmetry), a special case of Eq.\ (\ref{eq:deformation}) with 
only one non-zero $\la_m$, which renders the system non-integrable. In the following, 
we will discuss \emph{quadrupolar deformations} with $m=2$, which are the most 
energetically favorable of the multipole deformations, as well as the higher-order 
multipoles with $m\leq6$. Deformations with higher $m$ cost significantly more 
surface energy, and thus yield fewer stable configurations. The stability analysis of 
wires with 5- and 6-fold symmetry reveals only a few barely stable configurations 
(besides the axisymmetric wires discussed above), whereas for $m>6$ no stable 
deformed wires were found at all in the temperature range considered.

Figure \ref{fig:Stabdia:m2}(a) shows the stable configurations [thick curves] for 
quadrupolar Na wires at temperature $T=0.03\, T_\rho$.  These are determined by the 
intersection of the stationary curves, 
$\Omega^{(1)}(\bar{\rho},\bar{\lambda}_2)|_{\cal N}=0$, and the convex regions, 
$\Omega^{(2)}(\bar{\rho},\bar{\lambda}_2)|_{\cal N}>0$. Also shown are the thresholds 
to open new conducting channels, i.e., the energy eigenvalues of the two-dimensional 
billiard comprising the cross section of the wire.  Note that the stable structures 
lie in the gaps of the spectrum of channel thresholds, as is the case for 
axisymmetric wires.\cite{Kassubek01}

Combining results for all temperatures yields the stability diagram shown in Fig.\ 
\ref{fig:Stabdia:m2}(b). Note that the maximum deformation of the stable structures 
decreases strongly with increasing conductance. Nanowires with {\em highly-deformed 
cross sections are only stable at small conductance}, unlike the magic cylinders, 
which are predicted to occur for arbitrarily large conductance (until the transition 
to crystalline structures at sufficiently large radius).

\begin{table*}[ht]  
\begin{tabular}{|c||c|c|c|c||c|c|c||c|c|c|}
    \colrule
    $G$ & \multicolumn{4}{c||}{Ellipse} &
      \multicolumn{3}{c||}{$\lambda_2\cos(2\varphi)$} &
      \multicolumn{3}{c|}{$\lambda_2\cos(2\varphi)+\lambda_4\cos(4\varphi)$} \\
      $\left[G_0\right]$ &
      $\varepsilon$ &
      $\;\la_2$ &
      $\;\la_4\;$ &
      $T_{\rm{max}}/T_{\rho}$ &
      $\varepsilon$ &
      $\;\la_2$ &
      $T_{\rm{max}}/T_{\rho}$ &
      $\;\la_2\;$ &
      $\;\la_4\;$ &
      $T_{\rm{max}}/T_{\rho}$
\\
\colrule
  2 & 1.65 & 0.24 & 0.0440 & 0.40 & 1.7  & 0.26 & 0.50 & 0.25 & -0.04       & 0.50 \\
  5 & 1.32 & 0.14 & 0.0140 & 0.44 & 1.32 & 0.14 & 0.49 & 0.15 & -0.03/+0.08 & 0.49/0.23 \\
  9 & 1.24 & 0.11 & 0.0086 & 0.50 & 1.22 & 0.10 & 0.50 & 0.10 & 0.03/-0.03   & 0.48/0.44 \\
 29 & 1.14 & 0.07 & 0.0032 & 0.54 & 1.13 & 0.06 & 0.54 & 0.07/0.05 & 0.008/-0.015 & 0.44/0.44 \\
 59 & 1.09 & 0.04 & 0.0014 & 0.50 & 1.11 & 0.05 & 0.49 & 0.04 & -0.01        & 0.49 \\
 72 & 1.08 & 0.04 & 0.0011 & 0.40 & 1.08 & 0.04 & 0.39 & 0.04 & 0.009       & 0.36 \\
\colrule
\end{tabular}
   \caption{ Comparison of the most stable 
   deformed wires with elliptic (columns 2--5), quadrupolar (columns
   6--8), and  more general (columns 9--11) cross sections. The
   first column gives the quantized conductance of the corresponding
   wire. For the elliptic and quadrupolar wires, both the aspect
   ratio and the value of the deformation parameters (obtained from
   Eq.\ (\ref{eq:expandellipse}) for the former) are given. The
   maximum temperature of stability $T_{max}$ is given for each wire.
   The values for the elliptic wires are taken from Ref.\
   \onlinecite{Urban04}. Note that multiple values
   indicate multiple stable wires with the same conductance.
   In all cases the surface tension
   was set to $0.22\,\rm{N/m}$, corresponding to Na.\cite{Perdew91}
   }\label{tab:m24}
\end{table*}

The most stable wires are found at the same conductance values, and with similar 
semi-axis ratios, as the stable elliptical wires, discussed in Ref.\ 
\onlinecite{Urban04}. This is not surprising, since the ellipse can be approximated 
to leading order by a quadrupole,
\begin{eqnarray}
\label{eq:expandellipse}
    R_{\rm{ell}}(\varphi)&\approx&(1-\kappa^2)+
    (2\kappa+7\kappa^3)\cos(2\varphi)
    +3\kappa^2\cos(4\varphi)\nonumber\\
    &&+5\kappa^3\cos(6\varphi)+\cO(\kappa^4),
\end{eqnarray}
where $\kappa\equiv\frac{1}{4}(\varepsilon^2-1)/(\varepsilon^2+1)$ is given in terms 
of the aspect ratio $\varepsilon$. The maximum temperature $T_{max}$ up to which the 
wires are stable is in general of the same order as found for elliptical wires. 
Nevertheless, for the stable wire at $G=2G_0$, $T_{max}$ is 20 percent larger 
compared to the corresponding elliptical wire, indicating that a wire with two 
conducting channels will show a peanut-shaped cross section, rather than an elliptic 
one. Remarkably, this result, which was obtained by minimizing the electronic 
energy---and thus contains nothing of the atomic structure---is exactly what one 
would expect for a wire with two atoms in the cross section.

\begin{figure}[bt]  
    \begin{center}
         \includegraphics[width=0.95\columnwidth,draft=false]{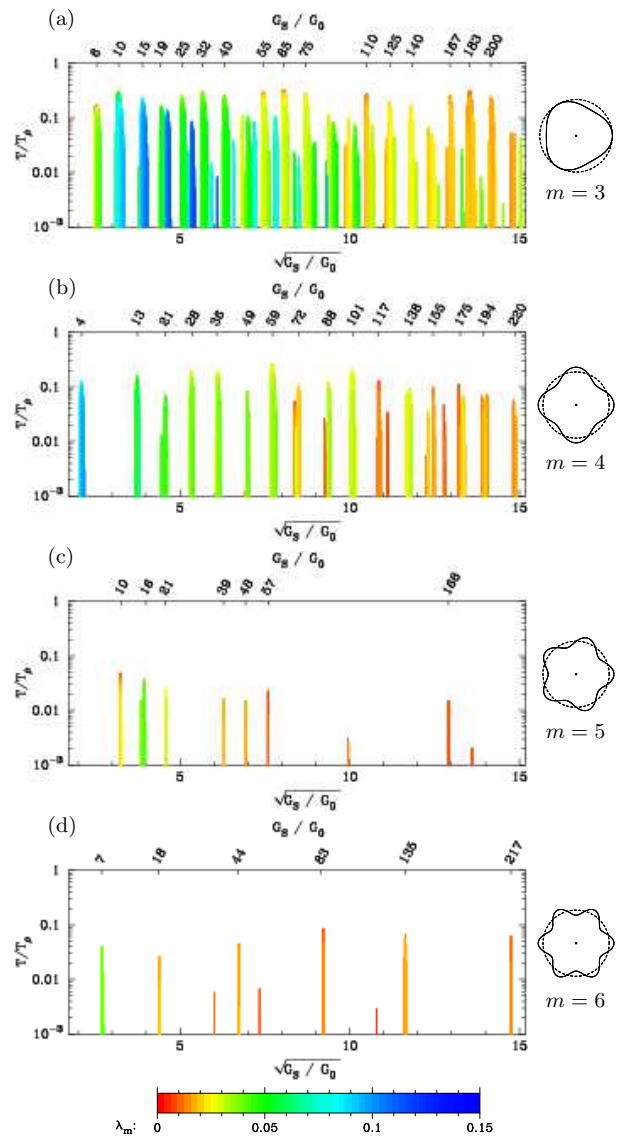}
    \end{center}
    \caption[]{(color online) Linear stability of
    $\cos(m\varphi)$-deformed wires for (a) $m=3$, (b) $m=4$, (c)
    $m=5$, and (d) $m=6$, as a function of temperature (in units of
    $T_{\rho}$). A logarithmic scale is used for the temperature
    since wires with $m\geq4$ are much less stable than those for
    $m\leq3$. A sketch of the cross section is shown on the right.
    The deformation parameter $\la_m$ is encoded via a gray(color)
    scale, common to all four diagrams. In the interest of clarity,
    only non-cylindrical wires (with $\la_m>0$) are shown. The
    surface tension was chosen to represent Na.}
    \label{fig:Stabdia:CosM}
\end{figure}

In order to shed light on the possible competition between integrable (elliptical) 
and non-integrable (quadrupolar) shapes, we analyzed the stability of wires with 
simultaneous $m=2$ and $m=4$ deformations, which can interpolate between elliptical 
and quadrupolar cross sections [cf.\ Eqs.\ (\ref{eq:deformation}) and 
(\ref{eq:expandellipse})]. The results are summarized in Table\ \ref{tab:m24}. For 
large deformations, the quadrupole is more stable, due to its lower surface energy, 
but no overall preference for a special geometry was found; stable wires with both 
positive and negative values of $\lambda_4$ exist. The values of $\lambda_4$ were 
found to be small, indicating that the addition of further deformations has little 
influence on the stability diagram. Remarkably, the wire with $G=2G_0$ is even closer 
to two touching cylinders when 4-fold deformations are added to the quadrupolar ones.

The linearly stable wires with 3-, 4-, 5-, and 6-fold symmetry are shown in Fig.\ 
\ref{fig:Stabdia:CosM} as a function of temperature. For clarity, the stable 
cylindrical wires are omitted, and a logarithmic scale is used for the temperature 
axis. The value of the deformation parameter $\la_m$ is encoded through a gray(color) 
scale, which is common to all stability diagrams shown in this figure, but is 
different than the one of Fig.\ \ref{fig:Stabdia:m2}. Compared to the quadrupolar 
wires, the number of stable configurations, the maximum temperature of stability, and 
the size of the deformations involved, all decrease rapidly with increasing order $m$ 
of the deformation. For $m>6$, we no longer find stable geometries in the temperature 
range considered. All this reflects the increase in surface energy with increasing 
order $m$ of the deformation.

From Fig.\ \ref{fig:Stabdia:CosM}(b), we can extract a series of stable wires with 
four-fold symmetry, for which the maximum deformation of $\la_4\sim 0.1$ is found at 
a conductance of $G=4\,G_0$. Although this geometry is expected to be far less stable 
than the neighboring cylindrical and quadrupolar wires, it has nevertheless likely 
been observed as a shoulder in Na conductance histograms.\cite{Urban04b}

\section{Material dependence}\label{sec:materialdependence}

Within the NFEM there is only one parameter entering the calculation apart from the 
contact geometry: the Fermi energy $E_F$, which is material dependent and in general 
well known (see Tab.\ \ref{tab.sigma.gamma}). Nevertheless, we have seen in the 
previous section that the energy cost of a deformation due to surface and curvature 
energy, which can vary significantly for different materials, plays a crucial role in 
determining the stability of a nanowire. Using the NFEM a priori implies the 
macroscopic free energy density $\omega=2E_Fk_F^3/15\pi^2$, the macroscopic surface 
energy $\sigmaS=E_Fk_F^2/16\pi$, and the macroscopic curvature energy 
$\gamma_s=2E_Fk_F/9\pi^2$. When drawing conclusions for metals having surface 
tensions and curvature energies that are rather different from these values, we have 
to think of an appropriate way to include these material-specific properties in our 
calculation.

Other authors have calculated the surface energy using a free-electron model with a 
rectangular confining potential of finite height,\cite{Huang49,Huntington51} 
depending on the electron work function, without being able to reproduce the values 
found in experiments. The agreement does not improve even when using a 
self-consistent confining potential.\cite{Huntington51} When working with a 
free-electron model, contributions of correlation and exchange energy are not 
included, which are found to play an essential role for a correct treatment of the 
surface energy. A discussion of the calculation of the surface energy of a jellium 
metal beyond the free-electron model can be found in Ref.\ \onlinecite{Mahan75}.

A convenient way of modeling the material properties without losing the pleasant 
features of the NFEM is via the implementation of the constraint 
(\ref{eq:constraint}) on the deformation, which interpolates between volume 
conservation ($\eta=\xi=0$) and treating the semiclassical expectation value for the 
charge $N_\weyl$ as an invariant ($\eta=3\pi/8$, $\xi=1$).

Consider the grand canonical potential of a free-electron gas confined within a given 
geometry by hard-wall boundaries, as given by Eq.\ (\ref{eq:OmegaWeyl}). The change 
in energy due to a deformation is
\begin{eqnarray}
    \Delta\Omega&=&
    -\omega\,\Delta\cV
    +\sigmaS\,\Delta\cS
    -\gamma_s\,\Delta\cC
    + \Delta[\delta \Omega]
\nonumber 
    \\
    &=&-\frac{\omega}{k_F^3}\Delta\cN
    +\Big(\sigmaS-
    \frac{\omega}{k_F}\eta\Big)\Delta\cS
\\
    & & -\Big(\gamma_s-\frac{\omega}{k_F^2}\xi\Big)\Delta\cC
    + \Delta[\delta \Omega],
    \nonumber
\end{eqnarray}
where we have used the constraint (\ref{eq:constraint}) to eliminate ${\cal V}$. Now 
the prefactors of the change in surface $\Delta\cS$ and the change in integrated mean 
curvature $\Delta\cC$ can be identified as effective surface tension and curvature 
energy, respectively. They can be adjusted to fit the material properties by 
appropriate choice\cite{Tdep} of the parameters $\eta$ and $\xi$ (see Tab.\ 
\ref{tab.sigma.gamma}). As noted previously, results in the long-wavelength limit are 
independent of $\xi$.

\begin{table}[tb]  
\begin{center}
\begin{tabular}{|l|c|c|c|c|c|c|}
\hline
    Element                   &   Li  &  Na   &  K    &  Cu   &   Ag  &  Au     \\
\hline
    $E_F$ [eV]                &  4.74 &  3.24 &  2.12 &  7.00 &  5.49 &  5.53   \\
    $k_F$ [nm$^{-1}$]         &  11.2 &   9.2 &   7.5 &  13.6 &  12.0 &  12.1   \\
\hline
    $\sigmaS$ [meV/\AA$^{2}$] & 27.2  & 13.6  & 7.58  & 93.3  & 64.9  & 78.5    \\
    $\sigmaS$ [$E_Fk_F^2$]    & 0.0046& 0.0050& 0.0064& 0.0072& 0.0082& 0.0097  \\
    $\eta$                    & 1.135 & 1.105 & 1.001 & 0.939 & 0.866 & 0.755   \\
\hline
    $\gamma_s$ [meV/\AA]      & 62.0  & 24.6  & 14.9  & 119   & 96.4  & 161     \\
    $\gamma_s$ [$E_Fk_F$]     & 0.0117& 0.0082& 0.0094& 0.0125& 0.0146& 0.0240  \\
    $\xi$                     & 0.802 & 1.06  & 0.971 & 0.741 & 0.583 & -0.111  \\
\hline
\end{tabular}
\end{center}
\vspace*{-0.2cm} \caption[]{Material parameters of several
  monovalent metals: Fermi energy $E_F$,\cite{Ashcroft-book}
  Fermi wavevector $k_F$,\cite{Ashcroft-book}
  surface tension $\sigmaS$, and curvature energy $\gamma_s$,\cite{Perdew91}
  along with the corresponding values of $\eta$ and $\xi$.
 } \label{tab.sigma.gamma}
\end{table}

The stability diagrams discussed in the previous sections all represent sodium 
nanowires. Results for other s-orbital metals are similar in respect to the number of 
stable configurations and the conductance of the wires. On the other hand, the 
deviations from axial symmetry and the relative stability of Jahn-Teller deformed 
wires is sensitive to the material specific surface tension $\sigmaS$ and Fermi 
temperature $T_F$. The relative stability of the highly deformed wires decreases with 
increasing surface tension $\sigmaS/E_Fk_F^2$, measured in intrinsic units, and this 
decrease becomes stronger with increasing order $m$ of the deformation. Therefore, 
for the simple metals under consideration (Tab.\ \ref{tab.sigma.gamma}), deformed Li 
wires have the highest, and Au wires have the lowest relative stability compared to 
cylinders of ``magic radii.''

However, concerning absolute stability, one must consider that the lifetime of a 
metastable nanowire also depends on the surface tension [cf.\ 
Eq.~(\ref{eq:universal})]. Thus, although the deformed wires have reduced relative 
stability in the noble metals, this effect is compensated by the greater absolute 
stability of these materials.

\begin{figure}[b]  
\begin{center}
    \includegraphics[width=\columnwidth,draft=false]{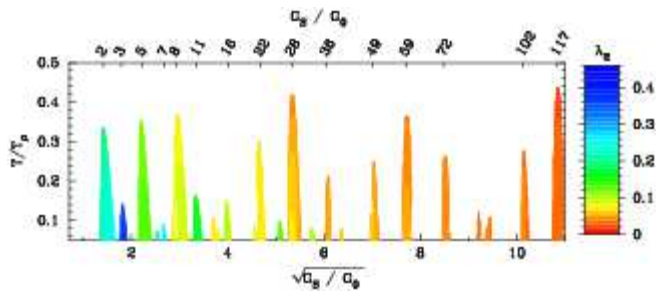}
\end{center}
\vspace*{-0.5cm} \caption[]{Linearly stable Au wires with
quadrupolar cross section. The deformation parameter $\lambda_2>0$ is
coded via a color(gray) scale.}\label{fig:Quadrupole:Au}
\end{figure}

In Fig.\ \ref{fig:Quadrupole:Au}, the linearly stable quadrupolar wires for gold are 
shown. The vertical scale was adjusted according to the arguments discussed above, in 
order to allow a comparison with the absolute stability for Na wires, as shown in 
Fig.\ \ref{fig:Stabdia:m2}.

\section{Summary and discussion}\label{sec:discussion}

In this article, we have performed a complete linear stability analysis of metal 
nanowires within the NFEM.  That is to say, we have examined the stability with 
respect to small, long-wavelength perturbations of metal nanostructures of arbitrary 
cross section, with translational symmetry along an axis perpendicular to the cross 
section.  We have considered both cross sections for which the electron motion is 
{\em integrable}, namely circles and ellipses, and nonintegrable cross sections, 
described by a multipole expansion (\ref{eq:deformation}).

This general stability analysis confirms the central conclusions of previous 
analyses, which were limited to circular\cite{Kassubek01,Urban03,Zhang03} and 
elliptical cross sections.\cite{Urban04,Urban04b} First, the existence of a sequence 
of ``magic'' cylindrical wires of exceptional stability, with electrical conductance 
$G/G_0=1$, 3, 6, 12, 17, 23, 34, 42, 51,..., was confirmed.  Second, the existence of 
a series of very stable deformed wires, with two-fold symmetric cross sections, which 
occur at the nodes of the cylindrical shell structure, at $G/G_0=2$, 5, 9, 29, 
59,..., was verified. For the thinnest of these wires ($G/G_0=2$, 5), which have the 
largest deformations, the quadrupolar cross section was found to be more stable than 
the ellipse, while the cross sections of the thicker wires in the sequence were found 
to be essentially elliptical. In addition, a number of new stable structures with 
quadrupolar, hexapolar, and octupolar cross sections have been identified.

Perhaps the most surprising result of our general linear stability analysis is the 
special role played by cylinders. Roughly 75\% of the principal structures observed 
in conductance histograms\cite{Yanson00} of alkali metals correspond to the magic 
cylinders.  The remaining 25\% are comprised of wires with elliptical/quadrupolar 
cross sections. Despite expanding the configuration space of allowed structures 
significantly in the present analysis, no new structures of comparable stability have 
been identified for the alkali or noble metals.  The role of symmetry in the 
stability of metal nanowires is thus fundamentally different from the case of atomic 
nuclei\cite{Magner97} or metal clusters,\cite{bulgac93,Schmidt99} where the vast 
majority of stable structures have broken symmetry.

The crucial difference between the stability of metal nanowires and metal clusters is 
not the shell effect, which is similar in both cases, but rather the surface energy, 
which favors the sphere, but abhors the cylinder.\cite{Chandrasekhar81} In order to 
understand the unique stability of metal nanocylinders, in contrast to other wire 
structures, it is useful to consider the semiclassical size of the competing surface 
and shell contributions:  From Eq.\ (\ref{eq:OmegaWeyl}), the surface energy scales 
as
\begin{equation}
    \label{eq:Omega_s}
    \frac{\Omega_s}{E_F k_F L} \sim {\cal O} (k_F\rho),
\end{equation}
where $\rho$ is the r.m.s.\ radius of the wire. On the other hand, Eq.\ 
(\ref{eq:TraceFormula}) implies that
\begin{equation}
    \label{eq:dOmega}
    \frac{\delta\Omega}{E_F k_F L} \sim {\cal O} (k_F \rho)^{-\nu},
\end{equation}
where $\nu=1$ for circular or elliptical cross sections, and $3/2 \geq \nu > 1$ for 
nonintegrable cross sections.  For a typical metal nanowire in the domain of validity 
of the NFEM, $k_F \rho \sim 10$ (i.e., $G_S/G_0 \sim 20$).  Thus, the term $\Omega_s$ 
which gives rise to the Rayleigh instability\cite{Kassubek01,Chandrasekhar81} is 
roughly two orders of magnitude larger than the shell correction $\delta\Omega$, 
which stabilizes the wire---even in the best-case scenario of an integrable cross 
section! (The volume contribution to $\Omega$, though larger still than $\Omega_s$, 
does not play a role in deformations that conserve the number of atoms.)  How then 
are we to understand the stability of metal nanowires?

Stability is not determined by the energy itself, but by the stationarity and 
convexity of the energy functional. Because $\delta \Omega$ is a rapidly oscillating 
function [cf.\ Eq.\ (\ref{eq:TraceFormula})], its derivatives are large in a 
semiclassical sense.\cite{Ullmo96} The first and second variations of $\delta \Omega$ 
scale as
\begin{equation}
    \label{eq:var_dOmega}
    \frac{\delta\Omega^{(1)}}{E_F k_F L} \sim {\cal O}(k_F \rho)^{1-\nu},
    \;\;\;\;
    \frac{\delta\Omega^{(2)}}{E_F k_F L} \sim {\cal O}(k_F \rho)^{2-\nu}.
\end{equation}
Thus, for the case of an integrable cross section ($\nu=1$), $\delta \Omega^{(2)}$ is 
of the same semiclassical size as $\Omega_s^{(2)}$.  However, $\delta \Omega^{(1)}$ 
is always semiclassically smaller than $\Omega_s^{(1)}$. The stationarity condition 
$\Omega^{(1)}=0$ is more difficult to satisfy than the convexity condition 
$\Omega^{(2)}>0$, which can be satisfied for any integrable cross section.

Cylinders are special because $\Omega^{(1)}=0$ by symmetry, so stability is 
determined solely by convexity, where surface and shell effects compete on an equal 
footing (cf. Fig.\ \ref{fig:StabCylinders}). In contrast, non-axisymmetric shapes can 
only occur for small $k_F \rho$.  More precisely, the maximum deviation from 
axisymmetry is a decreasing function of $k_F \rho$ (see Figs.\ \ref{fig:Stabdia:m2} 
and \ref{fig:Stabdia:CosM}). Both $\delta\Omega^{(1)}$ and $\Omega_s^{(1)}$ vanish as 
$\lambda\rightarrow 0$, but whereas $\delta\Omega^{(1)}$ is an oscillatory function 
of $\lambda$ [cf.\ Eq.\ (\ref{eq:TraceFormula})], $\Omega_s^{(1)}$ grows smoothly 
with increasing $\lambda$:
\begin{equation}
  \Omega_s^{(1)}/L \approx \sigma_s \lambda_m (m^2-1) 2 \pi \rho.
  \label{eq:omega_s_of_lambda}
\end{equation}
The stationarity condition $\Omega^{(1)}=0$ can therefore only be satisfied for
\begin{equation}
  |\lambda_m| < \frac{\mbox{const.}}{m^2-1}
    \left(\frac{E_F k_F^2}{\sigma_s}\right) (k_F \rho)^{-\nu}.
  \label{eq:max_lambda}
\end{equation}
Indeed, the decrease in the maximum deformations as a function of $k_F \rho$ in 
Figs.\ \ref{fig:Stabdia:m2}, \ref{fig:Stabdia:CosM}, and \ref{fig:Quadrupole:Au} is 
consistent with Eq.\ (\ref{eq:max_lambda}) with $3/2 \geq \nu >1$.

Thus cylinders are the only generically stable structures, and account for the 
principal peaks in conductance histograms of monovalent metals, while structures with 
broken symmetry are favored only at the nodes of the cylindrical supershell 
structure.

\section*{Acknowledgments}

We thank F.\ Kassubek, H.\ Tureci, and A.\ Douglas Stone for useful discussions. 
D.F.U.\ and C.A.S.\ acknowledge the Aspen Center of Physics, where the final stage of 
this project was carried out. This work was supported by the DFG and the EU Training 
Network DIENOW (D.F.U. and H.G.) and by NSF Grant Nos.\ 0312028 (J.B.\ and C.A.S.) 
and 0351964 (J.B.).

\appendix*

\section{Energy spectrum of a 2d-billiard} \label{app:EnergySpectrum}

The essential input for the stability analysis of a nanowire (Sec. \ref{sec:StabAna}) 
is the energy spectrum of a two-dimensional billiard for which the boundary (in polar 
coordinates $r,\varphi$) is given in terms of the radius function 
$R(\varphi;\rho,\La)\equiv R_\varphi$ of the wire, by the condition $r=R_\varphi$. 
This provides the transverse eigenenergies $E_{n}(\rho,\La)$ entering the expressions 
(\ref{eq:omega1N:m}) and (\ref{eq:stabcoefN:m}). It is helpful to introduce a change 
of coordinates, $\tilde{r}=r/R_\varphi$, which simplifies the boundary to a circle of 
unit radius. The Hamiltonian in the new coordinates $(\tilde{r},\varphi)$ can be 
written as $H=H^{(0)}+\delta H$, where 
$H^{(0)}=-R_{\varphi}^{-2}\triangle_{\tilde{r},\varphi}$ is the Hamiltonian for an 
axisymmetric problem and $\triangle_{\tilde{r},\varphi}$ is the Laplace operator. For 
simplicity of notation we use units in which $\hbar^2/2m_e=1$. On the other hand, 
$\delta H$ contains the terms due to deviations from axial symmetry and is found to 
read
\begin{eqnarray}
\label{eq:Hnew}
    \delta H
    &\!=\!&-
        \frac{{R_\varphi'}^2}{R_\varphi^4}\ddel{\tilde{r}}
        +2\frac{R_{\varphi}'}{R_{\varphi}^3}\frac{1}{\tilde{r}}\del{\tilde{r}}\del{\varphi}
    -\frac{2{R_{\varphi}'}^2\!-\!R_{\varphi}''R_{\varphi}}{R_{\varphi}^4}\frac{1}{\tilde{r}}\del{\tilde{r}},
    \nonumber\\
\end{eqnarray}
where a prime indicates differentiation with respect to $\varphi$. The energy 
spectrum can be obtained by numerical diagonalization of $H$ in an appropriate basis, 
that we choose to be
\begin{eqnarray}
\label{eq:basis}
    \Psi_{\mu\nu}(\tilde{r},\varphi)&=&\frac{1}{\sqrt{\pi}\cJ_{\mu+1}(\gamma_{\mu\nu})}\,
    e^{i\mu\varphi}\cJ_\mu(\gamma_{\mu\nu}\tilde{r}),
\end{eqnarray}
where the $\gamma_{\mu\nu}$ are the roots of the Bessel functions $\cJ_\mu$. We 
simplify the notation by using roman indices $j$ as multi-indices representing a set 
of quantum numbers $(\mu_j,\nu_j)$. Note that the states (\ref{eq:basis}), though 
normalized, are not orthogonal, since the scalar product in the new coordinates,
\begin{eqnarray}
\label{eq:scpnew} \left\langle {j}|{k}\right\rangle
&:=&\int_{0}^{2\pi}\!\!R_\varphi^2\,d\varphi\int_0^1 \tilde{r}\,d\tilde{r}\;
    \Psi_{j}^*(\tilde{r},\varphi)\Psi_{k}(\tilde{r},\varphi),\qquad
\end{eqnarray}
involves a factor $R_\varphi^2$. However, the $\{\Psi_{j}\}$ still form a basis and 
the energy spectrum can be obtained by solving a generalized eigenvalue problem. 
Defining the matrices $\cH_{jk}=\mel{j}{H}{k}$ and $\cB_{jk}=\scp{j}{k}$ we have to 
solve $\det\!\left[\cH-E\cB\right]=0$.
Taking matrix elements of $\delta H$ we obtain after some algebra,
\begin{eqnarray} \label{eq:Hdecomp}
    \left\langle {j}|\delta H|{k}\right\rangle&=&\bigg\langle
    {j}\bigg|\left(\gamma_{j}^2\delta_{\gamma_{j},\!\gamma_{k}}\!+
    \frac{\mu_{j}^2\gamma_{k}^2-\mu_{k}^2\gamma_{j}^2}{\gamma_{j}^2-\gamma_{k}^2}\,\frac{1}{\tilde{r}^2}\right)
    \frac{{R'_\varphi}^2}{R_\varphi^4}\bigg|\,{k}\bigg\rangle
\nonumber\\
&&+
    (\mu_{j}\!+\!\mu_{k})
    \bigg\langle
    {j}\bigg|\,\frac{iR'_\varphi}{R_\varphi^3}\frac{1}{\tilde{r}}\del{\tilde{r}}\bigg|\,{k}\bigg\rangle
\end{eqnarray}
where the terms could be simplified by making use of Bessel's equation.

We split the wave functions (\ref{eq:basis}) into angular and radial parts, 
$\phi_j(\varphi)$ and $\chi_j(\tilde{r})$, respectively, and introduce the following 
matrices of $\tilde{r}$- and $\varphi$-integrals
\begin{eqnarray}
    \cR^{[a]}_{jk}=\int_0^1d\tilde{r}\,\tilde{r}\chi_j^*\chi_k\,,\;\;
        &&
    \cI^{[a]}_{jk}=\int_0^{2\pi}\!d\varphi\,\phi_j^*\left(\frac{{{R_\varphi}'}}{{R_\varphi}}\right)^{\!\!2}\phi_k,
\nonumber\\
    \cR^{[b]}_{jk}=\int_0^1d\tilde{r}\,\frac{1}{\tilde{r}}\,\chi_j^*\,\chi_k\,,
        &&
    \cI^{[b]}_{jk}=\int_0^{2\pi}\!d\varphi\,\phi_j^*\left(\frac{i\,{{R_\varphi}'}}{{R_\varphi}}\right)\phi_k,
\nonumber\\
    \cR^{[c]}_{jk}=\int_0^1d\tilde{r}\chi_j^*\Del{\chi_k}{\tilde{r}}\,,\;\,
        &&
    \cI^{[c]}_{jk}=\int_0^{2\pi}\!d\varphi\left(\frac{{R_\varphi}}{\rho}\right)^2\phi_j^*\phi_k,
\nonumber\\
\label{eq:Def:Rmat:Imat}
\end{eqnarray}
Note that the $\cR^{[.]}$ do not depend on the deformation parameters $(\rho,\La)$ 
and that the $\cI^{[.]}$ only depend on $\La$. Decomposing the matrices $\cH$ and 
$\cB$ in terms of these newly defined matrices we find
\begin{eqnarray}
\label{eq:Hdecomposition}
    \lefteqn{\cH_{jk}=}\\
    &&\left\{\begin{array}{ll}
    \!\gamma_j^2\;\delta_{jk}+\Bigl(\gamma_j^2-\mu_j^2\cR^{[b]}_{jk}\Bigr)\cI^{[a]}_{jk};
    &
    \text{if}\,
    \gamma_j=\gamma_k,
    \\
    \!\frac{\mu_j^2\gamma_k^2-\mu_k^2\gamma_j^2}{\gamma_j^2-\gamma_k^2}\,\cR^{[b]}_{jk}\cI^{[a]}_{jk}
    +\,(\mu_j\!+\!\mu_k)\cR^{[c]}_{jk}\cI^{[b]}_{jk}
    &\text{if}\,
    \gamma_j\neq\gamma_k,
    \end{array}\right.
    \nonumber
    \\
\label{eq:Bdecomposition}
    \lefteqn{\cB_{jk}=\left\{\begin{array}{ll}
    \rho^2
    &
    \mbox{if}\quad j=k\,,
    \\
    \rho^2\cR^{[a]}_{jk}\cI^{[c]}_{jk}
    &
    \mbox{otherwise}\,.
    \end{array}\right.}
\end{eqnarray}
These results can now be implemented to solve the generalized eigenvalue problem 
numerically. Obviously, the larger the deviations from axial symmetry, the more basis 
functions have to be used in order to achieve good convergence.

\paragraph*{Expansion for small deformations of a cylinder:}
Moreover, Eqs.\ (\ref{eq:Hdecomposition}) and (\ref{eq:Bdecomposition}) also allow to 
study analytically the effect of a small deformation of a cylindrical wire on the 
transverse eigenenergies and the lifting of the degeneracies through the breaking of 
the rotational symmetry. Hereby we can give a proof of the relations 
(\ref{eq:DEandDDE:mu0}-\ref{eq:DDE:muNot0}) which ensure that the stability matrix of 
a cylindrical wire is always diagonal.

Let the initial shape be deformed according to the radius function 
(\ref{eq:deformation}). The deformation parameters are considered to be small so that 
we can expand the integrals $\cI^{[.]}$ up to second order in the $\la_m$. We find 
that they take non-zero values only if the quantum numbers $\mu_j$ and $\mu_k$ 
satisfy one of the following relations,
\begin{eqnarray}
\label{eq:dmu:conditionA}
    \mu_k-\mu_j &=& \pm\, n\cdot m\,,
\\
\label{eq:dmu:conditionB}
    |\mu_k-\mu_j| &=& |m\pm \bar{m}|\,,
\end{eqnarray}
where $n$ is an integer and $m$ and $\bar{m}$ are elements from the set $\bbmM$ of 
$m$-values appearing in the sum of Eq.\ (\ref{eq:deformation}). The results are 
summarized in Tab.\ \ref{tab:laExpansion}.

\begin{table}[htb]  
\begin{center}
    \begin{tabular}{|l|c|c|c|}
    \hline
      & $\cI^{[a]}_{jk}$ &
        $\cI^{[b]}_{jk}$ &
        $\cI^{[c]}_{jk}$ \\
    \hline
    $\mu_k\!-\!\mu_j=0$ &
        $\sum_m m^2\la_m^2/2$ & 
        $0$ &
        $1$ \\
    $\mu_k\!-\!\mu_j=\pm m$ &
        $\cO(\la_m^3)$ &
        $\pm\,m\,\la_m/2$   &
        $\la_m$             \\
    $\mu_k\!-\!\mu_j=\pm 2m$ &
        $-\,m^2\la_{m}^2/4$ &
        $\mp\,m\,\la_m^2/4$ &
        $\la_m^2/4$ \\
    $|\mu_k\!-\!\mu_j|=|m\!\pm\!\bar{m}|$  &
        $\mp\frac{mm'}{2}\la_m\la_{\bar{m}}$ &
        $\frac{1}{2}(\mu_j\!-\!\mu_k)\la_m\la_{\bar{m}}$&
        $\frac{1}{2}\la_m\la_{\bar{m}}$  \\
    otherwise &
        $\cO(\la_m^3)$ &
        $\cO(\la_m^3)$ &
        0 \\
    \hline
    \end{tabular}
    \caption[]{Expansion of the integrals $\cI^{[a]}$, $\cI^{[b]}$,
    and $\cI^{[c]}$ for small perturbations.
    Non-zero values are only found if one of the conditions given in
    the first column is satisfied.}
    \label{tab:laExpansion}
\end{center}
\end{table}

We can now straightforwardly derive the energy levels $E_j$ within perturbation 
theory of the generalized eigenvalue problem. Therefore we expand all the quantities 
entering the problem up to second order in the perturbation and solve order by order. 
For non-degenerate states ($\mu=0$) it is found that
\begin{eqnarray}
    \lefteqn{\rho^2E_{j}^{(1)}=\;\cH_{jj}^{(1)}-E_j^{(0)}\cB_{jj}^{(1)}}&&
\\
    \lefteqn{\rho^2E_j^{(2)}=}&&\\
&&
\cH_{jj}^{(2)}\!-E_j^{(1)}\cB_{jj}^{(1)}\!-E_j^{(0)}\cB_{jj}^{(2)}\!
        +\sum_{k\neq j}\frac{
    \bigl|\cH_{jk}^{(1)}\!-E^{(0)}_j\cB_{jk}^{(1)} \bigr|^2
    }{E_j^{(0)}-E_k^{(0)}}\nonumber
\end{eqnarray}
where the matrix $\cB$ in the equations above reflects the use of a non-orthogonal 
basis. From Eqs.\ (\ref{eq:Hdecomposition}) and (\ref{eq:Bdecomposition}) and the 
first row of Table\ \ref{tab:laExpansion} we deduce 
$\cH_{jj}^{(1)}=\cB_{jj}^{(1)}=\cB_{jj}^{(2)}=0$, so that $E_{j}^{(1)}=0$ and
\begin{eqnarray}
\label{eq:E-2ndOrderND}
    \lefteqn{\rho^2E_j^{(2)}=\;\cH_{jj}^{(2)}\;
        +\!\!\!\!\sum_{\genfrac{}{}{0pt}{}{k\neq j}{|\mu_k-\mu_j|=m}}\!\!\!\!
        \frac{\bigl|\cH_{jk}^{(1)}\!-E^{(0)}_j\cB_{jk}^{(1)} \bigr|^2}
             {E_j^{(0)}-E_k^{(0)}}}&&
\\
&=&\!
    \sum_m\lambda_m^2\!\left(\frac{\gamma_j^2\!-\!\mu_j^2\cR^{[b]}_{jj}}{2/m^2}\,
+\!\!\!\!\!\!\sum_{\genfrac{}{}{0pt}{}{k\neq j}{|\mu_k-\mu_j|=m}}\!\!\!\!\!\!\!
\frac{\left|\frac{\mu_j^2\!-\!\mu_k^2}{2}\cR^{[c]}_{jk}
        +\gamma_j^2\cR^{[a]}_{jk}\right|^2}{\gamma_j^2-\gamma_k^2}\right)\!\!.\nonumber
\end{eqnarray}
Since $E_j^{(2)}$ does not contain mixed terms $\propto\la_m\la_{\bar{m}}$ (with 
$m\neq\bar{m}$) we have hereby derived the relation~(\ref{eq:DEandDDE:mu0}).

The case $\mu>0$ is slightly more complicated as it requires degenerate perturbation 
theory: In first order we have to solve
\begin{eqnarray}
\label{eq:detGleichung}
    \left|
    \begin{pmatrix}
    \cH_{ii}^{(1)}\! & \cH_{ij}^{(1)}\\
    \cH_{ji}^{(1)}\! & \cH_{jj}^{(1)}
    \end{pmatrix}
    -E^{(0)}\!
    \begin{pmatrix}
    \cB_{ii}^{(1)}\! & \cB_{ij}^{(1)} \\
    \cB_{ji}^{(1)}\! & \cB_{jj}^{(1)}
    \end{pmatrix}
    -E^{(1)}\rho^2\unitMat\;
    \right|&=&0,\nonumber\\
\end{eqnarray}
where $\unitMat$ is the unit matrix, and $i=(\mu,\nu)$, and $j=(-\mu,\nu)$. Using the 
expansions given in Tab.\ \ref{tab:laExpansion}, we find
\begin{eqnarray}
\label{eq:E-1stOrderDeg}
    E_{i}^{(1)}&=&\left\{\begin{array}{ll}
    \pm\,(\gamma_i/\rho)^2&\mbox{if}\; 2\mu=m,\\
    0&\text{otherwise.}\end{array}\right.
    \qquad
\end{eqnarray}

The energy change in second order degenerate perturbation theory is found to read
\begin{eqnarray}
\label{eq:E-2ndOrderDeg}
    E_i^{(2)}\!\!&\!=\!&
        \cH_{ii}^{(2)}\!\!+\cC_{ii}
        \mp\left(\cH_{ij}^{(2)}\!\!-\!E_i^{(1)}\!\cB_{ij}^{(1)}\!\!-\!E_i^{(0)}\!\cB_{ij}^{(2)}\!\!+\cC_{ij}\!\right)
        \qquad\;
\end{eqnarray}
where we have used $\cB_{ii}^{(1)}=\cB_{ii}^{(2)}=0$, and the matrix
\begin{eqnarray}
    \cC_{ij}&=&\sum_{k\neq i,j}\frac{
    \bigl(\cH_{jk}^{(1)}-E^{(0)}_i\cB_{jk}^{(1)}\bigr)
    \bigl(\cH_{ik}^{(1)} -E^{(0)}_i\cB_{ik}^{(1)}\bigr)
    }{E_i^{(0)}-E_k^{(0)}}\;.\;\qquad
\end{eqnarray}
Note that the quantity $\cH_{ii}^{(2)}\!\!+\cC_{ii}$ appearing in Eq.\ 
(\ref{eq:E-2ndOrderDeg}) is equal to the full result (\ref{eq:E-2ndOrderND}) for the 
non-degenerate case. Therefore we conclude that relations (\ref{eq:DE:muNot0}) and 
(\ref{eq:DDE:muNot0}) hold.

\bibliography{SymBreaking}

\begin{thebibliography}{46}
\expandafter\ifx\csname natexlab\endcsname\relax\def\natexlab#1{#1}\fi
\expandafter\ifx\csname bibnamefont\endcsname\relax
  \def\bibnamefont#1{#1}\fi
\expandafter\ifx\csname bibfnamefont\endcsname\relax
  \def\bibfnamefont#1{#1}\fi
\expandafter\ifx\csname citenamefont\endcsname\relax
  \def\citenamefont#1{#1}\fi
\expandafter\ifx\csname url\endcsname\relax
  \def\url#1{\texttt{#1}}\fi
\expandafter\ifx\csname urlprefix\endcsname\relax\def\urlprefix{URL }\fi
\providecommand{\bibinfo}[2]{#2}
\providecommand{\eprint}[2][]{\url{#2}}

\bibitem[{\citenamefont{Agra{\"\i}t et~al.}(2003)\citenamefont{Agra{\"\i}t,
  Levy~Yeyati, and van Ruitenbeek}}]{Agrait03}
\bibinfo{author}{\bibfnamefont{N.}~\bibnamefont{Agra{\"\i}t}},
  \bibinfo{author}{\bibfnamefont{A.}~\bibnamefont{Levy~Yeyati}},
  \bibnamefont{and} \bibinfo{author}{\bibfnamefont{J.~M.} \bibnamefont{van
  Ruitenbeek}}, \bibinfo{journal}{Phys. Rep.} \textbf{\bibinfo{volume}{377}},
  \bibinfo{pages}{81} (\bibinfo{year}{2003}), \bibinfo{note}{and refs.
  therein.}

\bibitem[{\citenamefont{Rubio et~al.}(1996)\citenamefont{Rubio, Agra{\"\i}t,
  and Vieira}}]{Rubio96}
\bibinfo{author}{\bibfnamefont{G.}~\bibnamefont{Rubio}},
  \bibinfo{author}{\bibfnamefont{N.}~\bibnamefont{Agra{\"\i}t}},
  \bibnamefont{and} \bibinfo{author}{\bibfnamefont{S.}~\bibnamefont{Vieira}},
  \bibinfo{journal}{Phys. Rev. Lett.} \textbf{\bibinfo{volume}{76}},
  \bibinfo{pages}{2302} (\bibinfo{year}{1996}).

\bibitem[{\citenamefont{Untiedt et~al.}(1997)\citenamefont{Untiedt, Rubio,
  Vieira, and Agra{\"\i}t}}]{Untiedt97}
\bibinfo{author}{\bibfnamefont{C.}~\bibnamefont{Untiedt}},
  \bibinfo{author}{\bibfnamefont{G.}~\bibnamefont{Rubio}},
  \bibinfo{author}{\bibfnamefont{S.}~\bibnamefont{Vieira}}, \bibnamefont{and}
  \bibinfo{author}{\bibfnamefont{N.}~\bibnamefont{Agra{\"\i}t}},
  \bibinfo{journal}{Phys. Rev. B} \textbf{\bibinfo{volume}{56}},
  \bibinfo{pages}{2154} (\bibinfo{year}{1997}).

\bibitem[{\citenamefont{Kondo and Takayanagi}(1997)}]{Kondo97}
\bibinfo{author}{\bibfnamefont{Y.}~\bibnamefont{Kondo}} \bibnamefont{and}
  \bibinfo{author}{\bibfnamefont{K.}~\bibnamefont{Takayanagi}},
  \bibinfo{journal}{Phys. Rev. Lett.} \textbf{\bibinfo{volume}{79}},
  \bibinfo{pages}{3455} (\bibinfo{year}{1997}).

\bibitem[{\citenamefont{Rodrigues et~al.}(2000)\citenamefont{Rodrigues, Fuhrer,
  and Ugarte}}]{Rodrigues00}
\bibinfo{author}{\bibfnamefont{V.}~\bibnamefont{Rodrigues}},
  \bibinfo{author}{\bibfnamefont{T.}~\bibnamefont{Fuhrer}}, \bibnamefont{and}
  \bibinfo{author}{\bibfnamefont{D.}~\bibnamefont{Ugarte}},
  \bibinfo{journal}{Phys. Rev. Lett.} \textbf{\bibinfo{volume}{85}},
  \bibinfo{pages}{4124} (\bibinfo{year}{2000}).

\bibitem[{\citenamefont{Kondo and Takayanagi}(2000)}]{Kondo00}
\bibinfo{author}{\bibfnamefont{Y.}~\bibnamefont{Kondo}} \bibnamefont{and}
  \bibinfo{author}{\bibfnamefont{K.}~\bibnamefont{Takayanagi}},
  \bibinfo{journal}{Science} \textbf{\bibinfo{volume}{289}},
  \bibinfo{pages}{606} (\bibinfo{year}{2000}).

\bibitem[{\citenamefont{Yanson et~al.}(1999)\citenamefont{Yanson, Yanson, and
  van Ruitenbeek}}]{Yanson99}
\bibinfo{author}{\bibfnamefont{A.~I.} \bibnamefont{Yanson}},
  \bibinfo{author}{\bibfnamefont{I.~K.} \bibnamefont{Yanson}},
  \bibnamefont{and} \bibinfo{author}{\bibfnamefont{J.~M.} \bibnamefont{van
  Ruitenbeek}}, \bibinfo{journal}{Nature} \textbf{\bibinfo{volume}{400}},
  \bibinfo{pages}{144} (\bibinfo{year}{1999}).

\bibitem[{\citenamefont{Yanson et~al.}(2000)\citenamefont{Yanson, Yanson, and
  van Ruitenbeek}}]{Yanson00}
\bibinfo{author}{\bibfnamefont{A.~I.} \bibnamefont{Yanson}},
  \bibinfo{author}{\bibfnamefont{I.~K.} \bibnamefont{Yanson}},
  \bibnamefont{and} \bibinfo{author}{\bibfnamefont{J.~M.} \bibnamefont{van
  Ruitenbeek}}, \bibinfo{journal}{Phys. Rev. Lett.}
  \textbf{\bibinfo{volume}{84}}, \bibinfo{pages}{5832} (\bibinfo{year}{2000}).

\bibitem[{\citenamefont{Yanson et~al.}(2001)\citenamefont{Yanson, Yanson, and
  van Ruitenbeek}}]{Yanson01}
\bibinfo{author}{\bibfnamefont{A.~I.} \bibnamefont{Yanson}},
  \bibinfo{author}{\bibfnamefont{I.~K.} \bibnamefont{Yanson}},
  \bibnamefont{and} \bibinfo{author}{\bibfnamefont{J.~M.} \bibnamefont{van
  Ruitenbeek}}, \bibinfo{journal}{Fizika Nizkikh Temperatur}
  \textbf{\bibinfo{volume}{27}}, \bibinfo{pages}{1092} (\bibinfo{year}{2001}).

\bibitem[{\citenamefont{D{\'\i}az et~al.}(2003)\citenamefont{D{\'\i}az,
  Costa-Kr{\"a}mer, Medina, Hasmy, and Serena}}]{Diaz03}
\bibinfo{author}{\bibfnamefont{M.}~\bibnamefont{D{\'\i}az}},
  \bibinfo{author}{\bibfnamefont{J.~L.} \bibnamefont{Costa-Kr{\"a}mer}},
  \bibinfo{author}{\bibfnamefont{E.}~\bibnamefont{Medina}},
  \bibinfo{author}{\bibfnamefont{A.}~\bibnamefont{Hasmy}}, \bibnamefont{and}
  \bibinfo{author}{\bibfnamefont{P.~A.} \bibnamefont{Serena}},
  \bibinfo{journal}{Nanotech.} \textbf{\bibinfo{volume}{14}},
  \bibinfo{pages}{113} (\bibinfo{year}{2003}).

\bibitem[{\citenamefont{Mares et~al.}(2004)\citenamefont{Mares, Otte,
  Soukiassian, Smit, and van Ruitenbeek}}]{Mares04}
\bibinfo{author}{\bibfnamefont{A.~I.} \bibnamefont{Mares}},
  \bibinfo{author}{\bibfnamefont{A.~F.} \bibnamefont{Otte}},
  \bibinfo{author}{\bibfnamefont{L.~G.} \bibnamefont{Soukiassian}},
  \bibinfo{author}{\bibfnamefont{R.~H.~M.} \bibnamefont{Smit}},
  \bibnamefont{and} \bibinfo{author}{\bibfnamefont{J.~M.} \bibnamefont{van
  Ruitenbeek}}, \bibinfo{journal}{Phys. Rev. B} \textbf{\bibinfo{volume}{70}},
  \bibinfo{pages}{073401} (\bibinfo{year}{2004}).

\bibitem[{\citenamefont{Mares and van Ruitenbeek}(2005)}]{Mares05}
\bibinfo{author}{\bibfnamefont{A.~I.} \bibnamefont{Mares}} \bibnamefont{and}
  \bibinfo{author}{\bibfnamefont{J.~M.} \bibnamefont{van Ruitenbeek}},
  \bibinfo{journal}{Phys. Rev. B} \textbf{\bibinfo{volume}{72}},
  \bibinfo{pages}{205402} (\bibinfo{year}{2005}).

\bibitem[{\citenamefont{Zhang et~al.}(2003)\citenamefont{Zhang, Kassubek, and
  Stafford}}]{Zhang03}
\bibinfo{author}{\bibfnamefont{C.~H.} \bibnamefont{Zhang}},
  \bibinfo{author}{\bibfnamefont{F.}~\bibnamefont{Kassubek}}, \bibnamefont{and}
  \bibinfo{author}{\bibfnamefont{C.~A.} \bibnamefont{Stafford}},
  \bibinfo{journal}{Phys. Rev. B} \textbf{\bibinfo{volume}{68}},
  \bibinfo{pages}{165414} (\bibinfo{year}{2003}).

\bibitem[{\citenamefont{Chandrasekhar}(1981)}]{Chandrasekhar81}
\bibinfo{author}{\bibfnamefont{S.}~\bibnamefont{Chandrasekhar}},
  \emph{\bibinfo{title}{Hydrodynamic and Hydromagnetic Stability}}
  (\bibinfo{publisher}{Dover, New York}, \bibinfo{year}{1981}), pp.
  \bibinfo{pages}{515--74}.

\bibitem[{\citenamefont{de~Heer}(1993)}]{deHeer93}
\bibinfo{author}{\bibfnamefont{W.~A.} \bibnamefont{de~Heer}},
  \bibinfo{journal}{Rev. Mod. Phys.} \textbf{\bibinfo{volume}{65}},
  \bibinfo{pages}{611} (\bibinfo{year}{1993}).

\bibitem[{\citenamefont{Stafford et~al.}(1997)\citenamefont{Stafford,
  Baeriswyl, and B{\"u}rki}}]{Stafford97a}
\bibinfo{author}{\bibfnamefont{C.~A.} \bibnamefont{Stafford}},
  \bibinfo{author}{\bibfnamefont{D.}~\bibnamefont{Baeriswyl}},
  \bibnamefont{and}
  \bibinfo{author}{\bibfnamefont{J.}~\bibnamefont{B{\"u}rki}},
  \bibinfo{journal}{Phys. Rev. Lett.} \textbf{\bibinfo{volume}{79}},
  \bibinfo{pages}{2863} (\bibinfo{year}{1997}).

\bibitem[{\citenamefont{B{\"u}rki and Stafford}(2005)}]{Buerki05b}
\bibinfo{author}{\bibfnamefont{J.}~\bibnamefont{B{\"u}rki}} \bibnamefont{and}
  \bibinfo{author}{\bibfnamefont{C.~A.} \bibnamefont{Stafford}},
  \bibinfo{journal}{Applied Physics A} \textbf{\bibinfo{volume}{81}},
  \bibinfo{pages}{1519} (\bibinfo{year}{2005}).

\bibitem[{\citenamefont{Kassubek et~al.}(2001)\citenamefont{Kassubek, Stafford,
  Grabert, and Goldstein}}]{Kassubek01}
\bibinfo{author}{\bibfnamefont{F.}~\bibnamefont{Kassubek}},
  \bibinfo{author}{\bibfnamefont{C.~A.} \bibnamefont{Stafford}},
  \bibinfo{author}{\bibfnamefont{H.}~\bibnamefont{Grabert}}, \bibnamefont{and}
  \bibinfo{author}{\bibfnamefont{R.~E.} \bibnamefont{Goldstein}},
  \bibinfo{journal}{Nonlinearity} \textbf{\bibinfo{volume}{14}},
  \bibinfo{pages}{167} (\bibinfo{year}{2001}).

\bibitem[{\citenamefont{Urban and Grabert}(2003)}]{Urban03}
\bibinfo{author}{\bibfnamefont{D.~F.} \bibnamefont{Urban}} \bibnamefont{and}
  \bibinfo{author}{\bibfnamefont{H.}~\bibnamefont{Grabert}},
  \bibinfo{journal}{Phys. Rev. Lett.} \textbf{\bibinfo{volume}{91}},
  \bibinfo{pages}{256803} (\bibinfo{year}{2003}).

\bibitem[{\citenamefont{B{\"u}rki et~al.}(2003)\citenamefont{B{\"u}rki,
  Goldstein, and Stafford}}]{Buerki03}
\bibinfo{author}{\bibfnamefont{J.}~\bibnamefont{B{\"u}rki}},
  \bibinfo{author}{\bibfnamefont{R.~E.} \bibnamefont{Goldstein}},
  \bibnamefont{and} \bibinfo{author}{\bibfnamefont{C.~A.}
  \bibnamefont{Stafford}}, \bibinfo{journal}{Phys. Rev. Lett.}
  \textbf{\bibinfo{volume}{91}}, \bibinfo{pages}{254501}
  (\bibinfo{year}{2003}).

\bibitem[{\citenamefont{B{\"u}rki et~al.}(2005)\citenamefont{B{\"u}rki,
  Stafford, and Stein}}]{Buerki05}
\bibinfo{author}{\bibfnamefont{J.}~\bibnamefont{B{\"u}rki}},
  \bibinfo{author}{\bibfnamefont{C.~A.} \bibnamefont{Stafford}},
  \bibnamefont{and} \bibinfo{author}{\bibfnamefont{D.~L.} \bibnamefont{Stein}},
  \bibinfo{journal}{Phys. Rev. Lett.} \textbf{\bibinfo{volume}{95}},
  \bibinfo{pages}{090601} (\bibinfo{year}{2005}).

\bibitem[{\citenamefont{Bulgac and Lewenkopf}(1993)}]{bulgac93}
\bibinfo{author}{\bibfnamefont{A.}~\bibnamefont{Bulgac}} \bibnamefont{and}
  \bibinfo{author}{\bibfnamefont{C.}~\bibnamefont{Lewenkopf}},
  \bibinfo{journal}{Phys. Rev. Lett.} \textbf{\bibinfo{volume}{71}},
  \bibinfo{pages}{4130} (\bibinfo{year}{1993}).

\bibitem[{\citenamefont{Schmidt et~al.}(1999)\citenamefont{Schmidt, Ellert,
  Kronm\"uller, and Haberland}}]{Schmidt99}
\bibinfo{author}{\bibfnamefont{M.}~\bibnamefont{Schmidt}},
  \bibinfo{author}{\bibfnamefont{C.}~\bibnamefont{Ellert}},
  \bibinfo{author}{\bibfnamefont{W.}~\bibnamefont{Kronm\"uller}},
  \bibnamefont{and}
  \bibinfo{author}{\bibfnamefont{H.}~\bibnamefont{Haberland}},
  \bibinfo{journal}{Phys. Rev. B} \textbf{\bibinfo{volume}{59}},
  \bibinfo{pages}{10970} (\bibinfo{year}{1999}).

\bibitem[{\citenamefont{Urban et~al.}(2004{\natexlab{a}})\citenamefont{Urban,
  B{\"u}rki, Zhang, Stafford, and Grabert}}]{Urban04}
\bibinfo{author}{\bibfnamefont{D.~F.} \bibnamefont{Urban}},
  \bibinfo{author}{\bibfnamefont{J.}~\bibnamefont{B{\"u}rki}},
  \bibinfo{author}{\bibfnamefont{C.~H.} \bibnamefont{Zhang}},
  \bibinfo{author}{\bibfnamefont{C.~A.} \bibnamefont{Stafford}},
  \bibnamefont{and} \bibinfo{author}{\bibfnamefont{H.}~\bibnamefont{Grabert}},
  \bibinfo{journal}{Phys. Rev. Lett.} \textbf{\bibinfo{volume}{93}},
  \bibinfo{pages}{186403} (\bibinfo{year}{2004}{\natexlab{a}}).

\bibitem[{\citenamefont{Urban et~al.}(2004{\natexlab{b}})\citenamefont{Urban,
  B{\"u}rki, Yanson, Yanson, Stafford, van Ruitenbeek, and Grabert}}]{Urban04b}
\bibinfo{author}{\bibfnamefont{D.~F.} \bibnamefont{Urban}},
  \bibinfo{author}{\bibfnamefont{J.}~\bibnamefont{B{\"u}rki}},
  \bibinfo{author}{\bibfnamefont{A.~I.} \bibnamefont{Yanson}},
  \bibinfo{author}{\bibfnamefont{I.~K.} \bibnamefont{Yanson}},
  \bibinfo{author}{\bibfnamefont{C.~A.} \bibnamefont{Stafford}},
  \bibinfo{author}{\bibfnamefont{J.~M.} \bibnamefont{van Ruitenbeek}},
  \bibnamefont{and} \bibinfo{author}{\bibfnamefont{H.}~\bibnamefont{Grabert}},
  \bibinfo{journal}{Solid State Comm.} \textbf{\bibinfo{volume}{131}},
  \bibinfo{pages}{609} (\bibinfo{year}{2004}{\natexlab{b}}).

\bibitem[{\citenamefont{Brack and Bhaduri}(1997)}]{Brack97}
\bibinfo{author}{\bibfnamefont{M.}~\bibnamefont{Brack}} \bibnamefont{and}
  \bibinfo{author}{\bibfnamefont{R.~K.} \bibnamefont{Bhaduri}},
  \emph{\bibinfo{title}{Semiclassical Physics}}, vol.~\bibinfo{volume}{96} of
  \emph{\bibinfo{series}{Frontiers in Physics}}
  (\bibinfo{publisher}{Addison-Wesley, Reading, MA}, \bibinfo{year}{1997}).

\bibitem[{\citenamefont{Brack}(1993)}]{Brack93}
\bibinfo{author}{\bibfnamefont{M.}~\bibnamefont{Brack}}, \bibinfo{journal}{Rev.
  Mod. Phys.} \textbf{\bibinfo{volume}{65}}, \bibinfo{pages}{677}
  (\bibinfo{year}{1993}).

\bibitem[{\citenamefont{Kassubek et~al.}(1999)\citenamefont{Kassubek, Stafford,
  and Grabert}}]{Kassubek99}
\bibinfo{author}{\bibfnamefont{F.}~\bibnamefont{Kassubek}},
  \bibinfo{author}{\bibfnamefont{C.~A.} \bibnamefont{Stafford}},
  \bibnamefont{and} \bibinfo{author}{\bibfnamefont{H.}~\bibnamefont{Grabert}},
  \bibinfo{journal}{Phys. Rev. B} \textbf{\bibinfo{volume}{59}},
  \bibinfo{pages}{7560} (\bibinfo{year}{1999}).

\bibitem[{\citenamefont{Zhang et~al.}(2005)\citenamefont{Zhang, B{\"u}rki, and
  Stafford}}]{Zhang05}
\bibinfo{author}{\bibfnamefont{C.-H.} \bibnamefont{Zhang}},
  \bibinfo{author}{\bibfnamefont{J.}~\bibnamefont{B{\"u}rki}},
  \bibnamefont{and} \bibinfo{author}{\bibfnamefont{C.~A.}
  \bibnamefont{Stafford}}, \bibinfo{journal}{Phys. Rev. B}
  \textbf{\bibinfo{volume}{71}}, \bibinfo{pages}{235404}
  (\bibinfo{year}{2005}).

\bibitem[{\citenamefont{Garc{\'\i}a-Martin
  et~al.}(1996)\citenamefont{Garc{\'\i}a-Martin, Torres, and
  S{\'a}enz}}]{Garcia-martin96}
\bibinfo{author}{\bibfnamefont{A.}~\bibnamefont{Garc{\'\i}a-Martin}},
  \bibinfo{author}{\bibfnamefont{J.~A.} \bibnamefont{Torres}},
  \bibnamefont{and} \bibinfo{author}{\bibfnamefont{J.~J.}
  \bibnamefont{S{\'a}enz}}, \bibinfo{journal}{Phys. Rev. B}
  \textbf{\bibinfo{volume}{54}}, \bibinfo{pages}{13448} (\bibinfo{year}{1996}).

\bibitem[{\citenamefont{Martin}(1996)}]{Martin96}
\bibinfo{author}{\bibfnamefont{T.~P.} \bibnamefont{Martin}},
  \bibinfo{journal}{Phys. Rep.} \textbf{\bibinfo{volume}{273}},
  \bibinfo{pages}{199} (\bibinfo{year}{1996}).

\bibitem[{\citenamefont{Dashen et~al.}(1969)\citenamefont{Dashen, Ma, and
  Bernstein}}]{Dashen69}
\bibinfo{author}{\bibfnamefont{R.}~\bibnamefont{Dashen}},
  \bibinfo{author}{\bibfnamefont{S.-K.} \bibnamefont{Ma}}, \bibnamefont{and}
  \bibinfo{author}{\bibfnamefont{H.~J.} \bibnamefont{Bernstein}},
  \bibinfo{journal}{Phys. Rev.} \textbf{\bibinfo{volume}{187}},
  \bibinfo{pages}{345} (\bibinfo{year}{1969}).

\bibitem[{\citenamefont{Strutinsky}(1968)}]{Strutinsky68}
\bibinfo{author}{\bibfnamefont{V.~M.} \bibnamefont{Strutinsky}},
  \bibinfo{journal}{Nucl. Phys. A} \textbf{\bibinfo{volume}{122}},
  \bibinfo{pages}{1} (\bibinfo{year}{1968}).

\bibitem[{\citenamefont{Stafford et~al.}(1999)\citenamefont{Stafford, Kassubek,
  B{\"u}rki, and Grabert}}]{Stafford99}
\bibinfo{author}{\bibfnamefont{C.~A.} \bibnamefont{Stafford}},
  \bibinfo{author}{\bibfnamefont{F.}~\bibnamefont{Kassubek}},
  \bibinfo{author}{\bibfnamefont{J.}~\bibnamefont{B{\"u}rki}},
  \bibnamefont{and} \bibinfo{author}{\bibfnamefont{H.}~\bibnamefont{Grabert}},
  \bibinfo{journal}{Phys. Rev. Lett.} \textbf{\bibinfo{volume}{83}},
  \bibinfo{pages}{4836} (\bibinfo{year}{1999}).

\bibitem[{\citenamefont{Gutzwiller}(1990)}]{Gutzwiller90}
\bibinfo{author}{\bibfnamefont{M.~C.} \bibnamefont{Gutzwiller}},
  \emph{\bibinfo{title}{Chaos in Classical and Quantum Mechanics}}
  (\bibinfo{publisher}{Springer, New York}, \bibinfo{year}{1990}).

\bibitem[{\citenamefont{Gutzwiller}(1971)}]{Gutzwiller71}
\bibinfo{author}{\bibfnamefont{M.~C.} \bibnamefont{Gutzwiller}},
  \bibinfo{journal}{J. Math. Phys.} \textbf{\bibinfo{volume}{12}},
  \bibinfo{pages}{343} (\bibinfo{year}{1971}).

\bibitem[{\citenamefont{Stafford et~al.}(2001)\citenamefont{Stafford, Kassubek,
  and Grabert}}]{Stafford01}
\bibinfo{author}{\bibfnamefont{C.~A.} \bibnamefont{Stafford}},
  \bibinfo{author}{\bibfnamefont{F.}~\bibnamefont{Kassubek}}, \bibnamefont{and}
  \bibinfo{author}{\bibfnamefont{H.}~\bibnamefont{Grabert}},
  \bibinfo{journal}{Adv. Solid State Phys.} \textbf{\bibinfo{volume}{41}},
  \bibinfo{pages}{497} (\bibinfo{year}{2001}).

\bibitem[{\citenamefont{Urban et~al.}(2006)\citenamefont{Urban, Stafford, and
  Grabert}}]{Urban06}
\bibinfo{author}{\bibfnamefont{D.~F.} \bibnamefont{Urban}},
  \bibinfo{author}{\bibfnamefont{C.~A.} \bibnamefont{Stafford}},
  \bibnamefont{and} \bibinfo{author}{\bibfnamefont{H.}~\bibnamefont{Grabert}},
  \emph{\bibinfo{title}{Scaling theory of {CDW} in multi-channel metal
  nanowires}}, \bibinfo{howpublished}{preprint} (\bibinfo{year}{2006}).

\bibitem[{\citenamefont{Perdew et~al.}(1991)\citenamefont{Perdew, Wang, and
  Engel}}]{Perdew91}
\bibinfo{author}{\bibfnamefont{J.~P.} \bibnamefont{Perdew}},
  \bibinfo{author}{\bibfnamefont{Y.}~\bibnamefont{Wang}}, \bibnamefont{and}
  \bibinfo{author}{\bibfnamefont{E.}~\bibnamefont{Engel}},
  \bibinfo{journal}{Phys. Rev. Lett.} \textbf{\bibinfo{volume}{66}},
  \bibinfo{pages}{508} (\bibinfo{year}{1991}).

\bibitem[{\citenamefont{Huang and Wyllie}(1949)}]{Huang49}
\bibinfo{author}{\bibfnamefont{K.}~\bibnamefont{Huang}} \bibnamefont{and}
  \bibinfo{author}{\bibfnamefont{G.}~\bibnamefont{Wyllie}},
  \bibinfo{journal}{Proc. Phys. Soc. A} \textbf{\bibinfo{volume}{62}},
  \bibinfo{pages}{180} (\bibinfo{year}{1949}).

\bibitem[{\citenamefont{Huntington}(1951)}]{Huntington51}
\bibinfo{author}{\bibfnamefont{H.~B.} \bibnamefont{Huntington}},
  \bibinfo{journal}{Phys. Rev.} \textbf{\bibinfo{volume}{81}},
  \bibinfo{pages}{1035} (\bibinfo{year}{1951}).

\bibitem[{\citenamefont{Mahan}(1975)}]{Mahan75}
\bibinfo{author}{\bibfnamefont{G.~D.} \bibnamefont{Mahan}},
  \bibinfo{journal}{Phys. Rev. B} \textbf{\bibinfo{volume}{12}},
  \bibinfo{pages}{5585} (\bibinfo{year}{1975}).

\bibitem[{Tde()}]{Tdep}
\bibinfo{note}{Note that surface tension and curvature energy experimentally
  depend on temperature. This can be accounted for by making $\eta$ and $\xi$
  temperature dependent, but does not significantly alter the results.}

\bibitem[{\citenamefont{Ashcroft and Mermin}(1976)}]{Ashcroft-book}
\bibinfo{author}{\bibfnamefont{N.~W.} \bibnamefont{Ashcroft}} \bibnamefont{and}
  \bibinfo{author}{\bibfnamefont{N.~D.} \bibnamefont{Mermin}},
  \emph{\bibinfo{title}{Solid State Physics}} (\bibinfo{publisher}{Saunders
  College Publishing}, \bibinfo{year}{1976}).

\bibitem[{\citenamefont{Magner et~al.}(1997)\citenamefont{Magner, Fedotkin,
  Ivanyuk, Meier, Brack, Reimann, and Koizumi}}]{Magner97}
\bibinfo{author}{\bibfnamefont{A.~G.} \bibnamefont{Magner}},
  \bibinfo{author}{\bibfnamefont{S.~N.} \bibnamefont{Fedotkin}},
  \bibinfo{author}{\bibfnamefont{F.~A.} \bibnamefont{Ivanyuk}},
  \bibinfo{author}{\bibfnamefont{P.}~\bibnamefont{Meier}},
  \bibinfo{author}{\bibfnamefont{M.}~\bibnamefont{Brack}},
  \bibinfo{author}{\bibfnamefont{S.~M.} \bibnamefont{Reimann}},
  \bibnamefont{and} \bibinfo{author}{\bibfnamefont{H.}~\bibnamefont{Koizumi}},
  \bibinfo{journal}{Ann. Phys. (Leipzig)} \textbf{\bibinfo{volume}{6}},
  \bibinfo{pages}{555} (\bibinfo{year}{1997}).

\bibitem[{\citenamefont{Richter et~al.}(1996)\citenamefont{Richter, Ullmo, and
  Jalabert}}]{Ullmo96}
\bibinfo{author}{\bibfnamefont{K.}~\bibnamefont{Richter}},
  \bibinfo{author}{\bibfnamefont{D.}~\bibnamefont{Ullmo}}, \bibnamefont{and}
  \bibinfo{author}{\bibfnamefont{R.~A.} \bibnamefont{Jalabert}},
  \bibinfo{journal}{Phys. Rep.} \textbf{\bibinfo{volume}{276}},
  \bibinfo{pages}{1} (\bibinfo{year}{1996}).

\end{thebibliography}

\end{document}